\documentclass[%
 reprint,
nofootinbib,
 amsmath,amssymb,
 aps, showkeys,
]{revtex4-2}

\usepackage{graphicx}
\usepackage{dcolumn}
\usepackage{bm}
\usepackage{txfonts}
\usepackage{mathtools}
\usepackage[colorlinks=true, allcolors=blue]{hyperref}
\usepackage{esint}
\usepackage{siunitx}
\usepackage{float}
\usepackage{comment}

\newcommand\myeq{\stackrel{\mathclap{\tiny\mbox{BH}}}{=}}

\begin{document}

\preprint{APS/123-QED}

\title{Black Hole Horizons as Patternless Binary Messages\\and Markers of Dimensionality}

\author{Szymon Łukaszyk}
 \email{szymon@patent.pl}
\affiliation{%
 Łukaszyk Patent Attorneys, ul. Głowackiego 8, 40-052 Katowice, Poland
}%


\begin{abstract}
This study aims to reconcile quantum theory with the universality of the speed of light in vacuum and its implications on relativity through an information-theoretic approach. We introduce the concepts of a holographic sphere and variational potential. Entropy variation expressed in terms of the information capacity of this sphere results in the concept of binary potential in units of negative, squared speed of light in vacuum. Accordingly, the event horizon is a fundamental holographic sphere in thermodynamic equilibrium with only one exterior side: a noncompressible binary message that maximizes Shannon entropy. Therefore, the Jordan-Brouwer separation theorem and generalized Stokes theorem do not hold for black holes. We introduce the concept of inertial potential and demonstrate its equivalence to the variational potential, which ensures that any inertial acceleration represents a nonequilibrium thermodynamic condition. We introduce the concept of the complementary time period and relate it with the classical time period through integral powers of the imaginary unit to formulate the notions of unobservable velocity and acceleration, which are perpendicular and tangential to the holographic sphere, respectively, and bound with the observable velocity and acceleration based on Pythagorean relations. We further discuss certain dynamics scenarios between the two masses. The concept of black hole informationless emission is introduced as a complement to informationless Bekenstein absorption and extended to arbitrary wavelengths. Black hole quantum statistics with degeneracy interpreted as the number of Planck areas on the event horizon are discussed. The study concludes that holographic screens and equipotential surfaces are spherical equivalents, and every observer is a sphere in nonequilibrium thermodynamic condition. Lastly, we propose a solution to the black hole information paradox.
\end{abstract}


\maketitle


\section{Introduction}\label{sec:1}

An uncharged, nonrotating (Schwarzschild) black hole (BH) can be observed as a 2-sphere; however, one cannot state anything regarding its interior \cite{1}, which is equivalent to the fact that it does not have any interior. However, mathematical 2-spheres possess interiors. In general, any $(n-1)$-dimensional topological sphere in $\mathbb{R}^{n}$ forms a common boundary of two connected components: bounded interior and unbounded exterior, as asserted by the Jordan-Brouwer separation theorem. However, this theorem does not hold for BHs thereby disapproving the correctness of our intuitive concept of the physical space represented as $\mathbb{R}^{3}$ (or $\mathbb{R}^{4}$). Additionally, the generalized Stokes theorem does not hold for BHs. A differential form over a BH event horizon does not equal the integral of its exterior derivative over the BH interior, considering there is no such thing as a BH interior.

A BH elevates John Archibald Wheeler's ''\emph{it from bit}'' conclusion, which states that there is no such thing as space or time or spacetime continuum \cite{2} beyond the microscopic level. This causes us to research nature as a vertex-labeled graph (graph of nature) with certain intrinsic properties reflecting the $2^{\text {nd}}$ law of thermodynamics. The primordial Big Bang singularity (the first point or vertex) expanded this graph of nature into dimensionalities and not into a four-dimensional spacetime \cite{3}.

The combinatorial proof of the $H$-theorem derived by Ludwig Boltzmann in 1877 \cite{4} introduced energy quantization, which resulted in the development of quantum theory \cite{5} and created the measurement problem that demands an interpretation. To date, no consensus has been achieved in the scientific community regarding the universal validity of any particular interpretation of the measurement problem \cite{6}.

The author is convinced that life is an explanation of the measurement problem performed \emph{hic et nunc} by any particular observer. The measurement yields rational (as rational numbers) and classical (as bits, not qubits) information that is twofold, considering it relates both to the \emph{without} [exterior] and \emph{within} [interior] of things. Quoting Pierre Teilhard de Chardin, ''In the eyes of the physicist, nothing exists legitimately, at least up to now, except the \emph{without} of things. The same intellectual attitude is still permissible in the bacteriologist, whose cultures (apart from some substantial difficulties) are treated as laboratory reagents. But it is already more difficult in the realm of plants. It tends to become a gamble in the case of a biologist studying the behavior of insects or coelenterates. It seems merely futile with regard to the vertebrates. Finally, it breaks down completely with man, in whom the existence of a \emph{within} can no longer be evaded, because it is the object of a direct intuition and the substance of all knowledge'' \cite{7}.

Albert Einstein believed \cite{8} that ''the experience of the \emph{now} means something special for man, something essentially different from the past and the future, but that this important difference does not and cannot occur within physics. That this experience cannot be grasped by science seemed to him a matter of painful but inevitable resignation. (...) there is something essential about the \emph{now} which is just outside of the realm of science''. The author strongly disagrees with this statement: the 3-dimensional universe emerges out of an uncountably infinite-dimensional graph of nature during observer-dependent measurements. The observer-independent measurements simply do not exist \cite{9,10,11}. A local time associated with each observer in special relativity is reversible. Nonetheless, we can now investigate time in other conceptual contexts \cite{12}, especially as an emergent, entropic quantity\footnote{''Einstein imagined himself riding a photon. But (...) we are too heavy to ride photons or electrons. We cannot possibly replace such airy beings, identify ourselves with them, and describe what they would think, if they were able to think, and what they would experience, if they were able to feel anything'' (\cite{12} p. 219).}.

The author conjectures that biological evolution is possible only in complex \cite{manin} dimension $n=3+0i$ with 3 spatial dimensions and imaginary time owing to the exotic $\mathbb{R}^{4}$ property of such a space \cite{3}, the dimensionality of which was exploited by biological evolution from among uncountably infinite other dimensionalities, including fractional ones. Particularly, the only real imaginary number is $0 i=0$. This zero is the \emph{nunc}, that is the moment that passes for every living biological cell, and it cannot be placed outside the realm of science \cite{8,12}. Therefore, the Lorentz transformations denote the rotations of a 4-ball of a fixed radius $R$, expressed as
\begin{equation}\label{1}
x^{2}+y^{2}+z^{2}+(i c t)^{2}=R^{2} .
\end{equation}
Rather than aiming to propose new physics, this research aims to seek a new meaning to the existing, experimentally verified physical theories, especially to the theory of relativity, which is grounded on the objective, unobservable, extra-instantaneous existence of 3D physical reality (\emph{the universe}), and hence, is not complete.

This paper is structured as follows. Sections \ref{sec:2} and \ref{sec:3} present the discretization in terms of Planck units and the discrete BHs, respectively. Section \ref{sec:4} introduces the concept of a BH informationless emission as complementary to informationless Bekenstein absorption and extends it to arbitrary wavelengths. Subsequently, the author's perspective on the entropic gravity derivation disclosed in \cite{14} along with a simplified model of the spherical universe with two masses is presented in Section \ref{sec:5}. Thereafter, the concept of the variational potential and the entropy variation sphere is proposed in Section \ref{sec:6} to display the equivalence of known entropic gravity equations. Section \ref{sec:7} presents the concept of the binary potential to exemplify that an event horizon is a binary message that maximizes Shannon entropy. The concept of inertial potential is introduced in Section \ref{sec:8}. The concept of complementary time period related to \emph{classical} time period by integral powers of the imaginary unit is introduced in Section \ref{sec:9}, which includes the kinematics of the two masses on a holographic sphere, and the significant radii of a constant mass. Section \ref{sec:10} summarizes the Bose-Einstein, Maxwell-Boltzmann, and Fermi-Dirac statistics with degeneracy interpreted as the number of Planck areas on a BH event horizon. A solution to BH information paradox is proposed in Section \ref{sec:11}. Section \ref{sec:12} concludes the findings of this study.

The author is of the opinion that the notion of charge and other already unified quantities should emerge from topological and entropic considerations, and hence, charged BHs were deliberately neglected from the considerations of this study.

Baroque derivations were used to support certain controversial claims that largely deviate from the mainstream understanding of physics.

Natural units were also deliberately neglected from the considerations of this study to avoid loss of clarity. The meaning of all seven constants of nature used in this study: $c$ (speed of light in vacuum), $e$ (the base of the natural logarithm), $G$ (the gravitational constant), $h$ (the Planck constant), $i$ (the imaginary unit), $k_{B}$ (the Boltzmann constant), and $\pi$ (the ratio of a circle's circumference to its diameter), should not be treated carelessly.

A two-dimensional Boolean lattice (a holographic screen), evolving with time, deduced from unitarity, entropy, and counting arguments has been proposed in \cite{13} to explain the failure of developing completely consistent mathematical models of quantum BHs. This research aims to pursue this idea further in an attempt to reconcile quantum theory with the universality of the speed of light in vacuum and its implications on relativity, in an information-theoretic approach.

\section{Discretizations}\label{sec:2}
Discretization in terms of Planck units was adopted in certain segments of this study. These ''natural units of measure'' introduced by Max Planck in 1899 ''are independent of special bodies or substances, thereby necessarily retaining their meaning for all times and for all civilizations, including extraterrestrial and non-human ones'' \cite{15}.

Where deemed appropriate, uppercase letters denote dimensional quantities and lowercase letters denote Planck units or their multipliers. Masses $M$ were considered as real multiplicities $m$ of the Planck mass $m_{P}$ without loss of generality (w.l.o.g.), expressed as
\begin{equation}\label{2}
M \doteq m m_{P} \quad m \in \mathbb{R}.
\end{equation}
Additionally, lengths were considered w.l.o.g. as the real multiplicities of the Planck length $\ell_{P}$. Particularly, sphericality was considered w.l.o.g. in terms of the diameter $d$ or radius $r$ multipliers
\begin{equation}\label{3}
D \doteq d \ell_{P} \quad R \doteq r \ell_{P} \quad d, r \in \mathbb{R}.
\end{equation}
However, the wavelengths were considered as
\begin{equation}\label{4}
\lambda \doteq l \ell_{P} \quad l \in \mathbb{R} \backslash\{(-1,1)\},
\end{equation}
considering the wavelength denotes the distance over which the shape of the wave repeats, and this distance must be at least equal to the Planck length. No physical models describe smaller lengths; and hence the open set $(-1,1)$ is forbidden.

Based on mass \eqref{2} and wavelength \eqref{4} discretizations, the relation between wavelength $l$ and mass $m$ multipliers can be derived in terms of the Compton wavelength, expressed as
\begin{equation}\label{5}
\lambda_{M} \doteq l \ell_{P}=\frac{h}{c m m_{P}} \Leftrightarrow l m=2 \pi.
\end{equation}
On constraining to $l \geq 1$, we get
\begin{equation}\label{6}
m \leq 2 \pi \text {, }
\end{equation}
which can be called the \emph{threshold of distinguishability}, considering it indicates that the Compton wavelengths of masses $M \geq 2 \pi m_{P}$ (\num{1.37E-7} kg) are smaller than the Planck length, which is physically impossible. Such masses no longer exhibit wave-particle duality and would not interfere with each other in the double-slit experiment. In principle, they are neither bosonic nor fermionic (nor anyonic) and are distinguishable in all instances. Evidently, this does not preclude the distinguishability of lighter masses.

The possible meaning of negative wavelengths, and diameters is discussed in Sections \ref{sec:4}, \ref{sec:8}, \ref{sec:10}, and in the Appendices of this paper. The complex wavelength multipliers $l$ expressed in wavelength discretization \eqref{4} are briefly discussed in Sections \ref{sec:4} and \ref{sec:10}. Overall, detailed investigations of these notions are beyond the scope of this study. We note in passing, that complex geodesic paths emerge in the presence of BH singularities \cite{fidkowski} and when studying entropic dynamics on curved statistical manifolds \cite{gassner}.

\section{Discrete Black Holes}\label{sec:3}
On expressing the mass and diameter of the BH w.l.o.g. using mass \eqref{2} and diameter \eqref{3} discretizations, the Schwarzschild diameter can be expressed as
\begin{equation}\label{7}
D_{BH}=\frac{4 G m_{BH}}{c^{2}} \sqrt{\frac{\hbar c}{G}}=4 m_{BH} \ell_{P},
\end{equation}
yielding a simple relation between the BH mass and the diameter or radius multipliers, given as
\begin{equation}\label{8}
m_{BH}=\frac{d_{BH}}{4}=\frac{r_{BH}}{2} .
\end{equation}
Therefore, the BH Compton wavelength \eqref{5} can be expressed as
\begin{equation}\label{9}
l_{BH}=\frac{2 \pi}{m_{BH}}=\frac{8 \pi}{d_{BH}} \quad\left(\lambdabar_{BH}=\frac{2 \ell_{P}^{2}}{R_{BH}}=\frac{4 \ell_{P}^{2}}{D_{BH}}\right),
\end{equation}
and constraining to $l \geq 1$, the BH \emph{threshold of distinguishability} \eqref{6} can be rewritten in terms of a BH diameter or radius multipliers as
\begin{equation}\label{10}
d_{BH} \leq 8 \pi, \quad \text { or } \quad r_{BH} \leq 4 \pi.
\end{equation}
Therefore, only the BHs with diameter (radius) multipliers below this threshold are adequately small to have Compton wavelengths \eqref{5} greater than the Planck length. As we shall see in Section \ref{sec:4}, BH threshold of distinguishability \eqref{10} is equal to a threshold of BH collapsibility \eqref{22}.

\begin{figure}[h]
\includegraphics[width=0.5\textwidth]{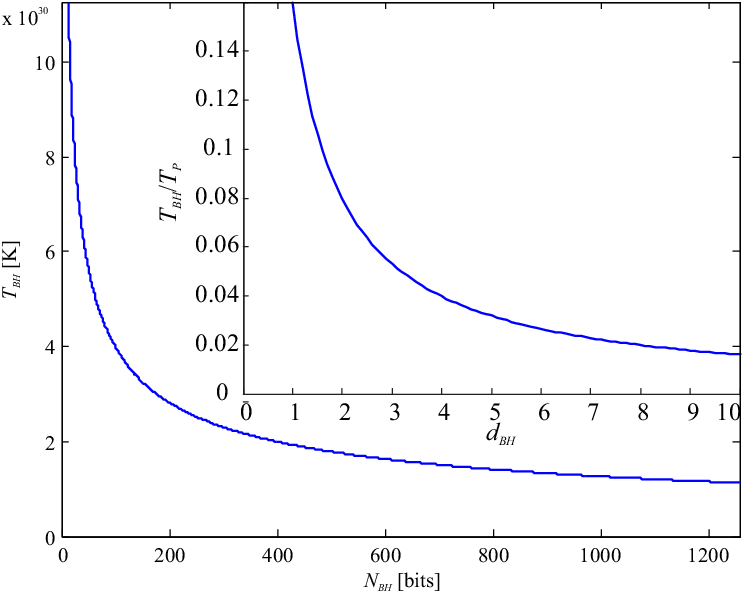}
\caption{\label{Fig_01} Black hole temperature $T_{BH}$ as a function of information capacity $N_{BH}$ and diameter multiplier $d_{BH}$.}
\end{figure}

Diameter discretization \eqref{3} used to determine the information capacity of an event horizon yields $N_{BH}=\pi D_{BH}^{2} / \ell_{P}^{2}=\pi d_{BH}^{2}$. Moreover, the BH surface gravity \eqref{82} substituted into the Hawking radiation formula \eqref{39} yields (cf. Appendix \ref{app:1}) the Hawking radiation temperature $T_{BH}$ as a function of diameter multiplier $d_{BH}$
\begin{equation}\label{11}
T_{BH}=\frac{1}{2 \pi d_{BH}} T_{P},
\end{equation}
as shown in Fig.~\ref{Fig_01}, where $T_{P}$ denotes the Planck temperature.

Given that no physical models define temperatures greater than the Planck temperature, the highest physically meaningful Big Bang temperature corresponds to a BH with $d_{BH}=1 /(2 \pi), N_{BH}=1 /(4 \pi)$, and $\left\lfloor N_{BH}\right\rfloor=0$. Furthermore, the natural diameter multiplier $d_{BH}$ can be related to the $2 \pi$-cycle: BH temperature \eqref{11} decreases by a factor of $2 \pi$ with every integer increment of $d$. At $d_{BH}=1$, BH is at a temperature $T_{BH}=T_P / 2 \pi$, which is the reduced Planck temperature, similar to the reduced Planck constant or reduced wavelength. The BH Hagedorn temperature of approximately \num{1.70E12} K - \emph{the boiling point of ordinary matter} - is achieved by the BH with $d_{BH} \approx \num{1.33E19}$, equivalent to $D_{BH}=\num{2.14E-16}$ m, which is approximately $1/4$ of a proton charge radius (\num{8.42E-16} m).

Table \ref{table1} summarizes certain examples of discrete BHs and displays the information capacity, number of bits, mass, and temperature of each hole as a function of its diameter multiplier $d_{BH}$.

\begin{table*}
\caption{\label{table1}Examples of discrete BHs.}
\begin{ruledtabular}
\begin{tabular}{|c|c|c|c|c|c|}
 $d_\text{BH}=D_{BH}/\ell_{P}$ & $N_{BH}=\pi d_{BH}^2$ & $\left\lfloor N_{BH}\right\rfloor$ & $M_{BH}/m_P=d_{BH}/4$ & $T_{BH}/T_P = 1/(2\pi d_{BH})$ & BH type/comments \\
 \hline
 $1/(2\pi)$ & $0.0796$ & $0$ & $0.0398$ & $1$ & The Big Bang at the Planck temperature \\
 $1/\pi$ & $0.3183$ & $0$ & $0.0796$ & $0.5$ & Half of the Planck temperature \\
 $1/\sqrt{\pi}$ & $1$ & $1$ & $0.1410$ & $0.2821$ & min $1$-bit, MB stat. $1^\text{st}$ singularity \\
 $\sqrt{\ln(4)/\pi}$ & $2\ln(2)$ & $1$ & $0.1661$ & $0.2396$ & Landauer $1^\text{st}$ bound ($M_{BH}c^2=T_{BH}k_B\ln(2)$) \\
 $\sqrt{2/\pi}$ & $2$ & $2$ & $0.1995$ & $0.1995$ & min $2$-bit, FD stat. $1^\text{st}$ singularity \\
 $2\sqrt{\ln(2)/\pi}$ & $4\ln(2)$ & $2$ & $0.2349$ & $0.1694$ & Landauer $2^\text{nd}$ bound (BH entropy) \\
 $1$ & $\pi$ & $3$ & $0.25$ & $0.1592$ & $\pi$-bit, surface gravity equals the Planck acc.\\
 $\sqrt{4/\pi}$ & $4$ & $4$ & $0.2821$ & $0.1410$ & min $4$-bit, one unit of entropy\\
 $\sqrt{2}$ & $2\pi$ & $6$ & $0.3536$ & $0.1125$ & the imaginary Planck time at the BH horizon\\
 $2$ & $4\pi$ & $12$ & $0.5$ & $0.0796$ & $12$-bit, HUP yields $R_{BH}=\ell_P$\\
 $4$ & $16\pi$ & $50$ & $1$ & $0.0398$ & $M_{BH}=m_P$\\
 $2\pi$ & $4\pi^3$ & $124$ & $1.5708$ & $0.0253$ & Margolus-Levitin theorem threshold\\
 $4\sqrt{\pi}$ & $16\pi^2$ & $157$ & $1.7725$ & $0.0224$ & Threshold of informationless stability\\
 $8\pi$ & $64\pi^3$ & $1984$ & $2\pi$ & $1/(16\pi^2)$ & $M_{BH}=2\pi m_P$, thr. of distinguishability\\
 \num{1.52E45} & \num{7.24E90} & \num{7.24E90} & \num{3.80E44} & \num{1.05E-46} & Sagittarius A$^*$
\end{tabular}
\end{ruledtabular}
\end{table*}

\begin{figure}[h]
\includegraphics[width=0.5\textwidth]{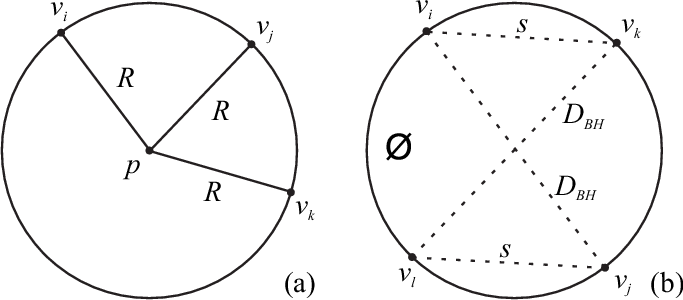}
\caption{\label{Fig_02} Surface of (a) $n$-ball, (b) BH.}
\end{figure}

Any vertex $v$ on the surface of an $n$-ball is related to a certain vertex $p$, which is the center of this $n$-ball by the same parameter called radius $R$. Formally, for a set of vertices $V$ with a generalized distance function $\mu \geq 0$ (cf. Appendix \ref{app:5}), the surface $A$ of the $n$-ball is defined as
\begin{equation}\label{12}
A(p)=\{v \in V \mid \mu(v, p)=R\} .
\end{equation}
However, this definition does not hold for BHs in any dimension $n$. As BHs do not have interiors, no vertex $p$ of the graph of nature can serve as the BH center. Thus, the term Schwarzschild \emph{radius} is a misnomer. To define a BH as a centerless $(n-1)$-sphere, the pairs of vertices on its surface must be related to each other through the same diameter $D_{BH}$. Therefore, the event horizon $A_{BH}$ of a BH can be defined as
\begin{equation}\label{13}
A_{BH}(p \in \varnothing)=\left\{\begin{array}{l}
\forall v_{i} \in V \; \exists! v_{j} \in V \; \cap \\ \forall v_{k} \in V \; \exists! v_{l} \in V \\
\mu\left(v_{i}, v_{j}\right)=\mu\left(v_{k}, v_{l}\right)=D_{BH} \; \cap \\ 
\mu\left(v_{i}, v_{k}\right)=\mu\left(v_{j}, v_{l}\right)
\end{array}\right\} .
\end{equation}
Each vertex of the event horizon is associated with precisely one vertex based on the relation of the diameter $D_{BH}$, and each pair of vertices non-associated with the relation of the diameter $D_{BH}$ is associated with each other based on the distance relation equal to that of their $D_{BH}$ partners. Only the latter condition is required for a BH to be perceived as \emph{spherical}. The definition \eqref{13} requires at least four vertices defining two diameters, as shown in Fig. \ref{Fig_02}(b) (the lines in the figure do not indicate line segments passing across a BH interior).

\begin{figure}[ht]
\includegraphics[width=0.4\textwidth]{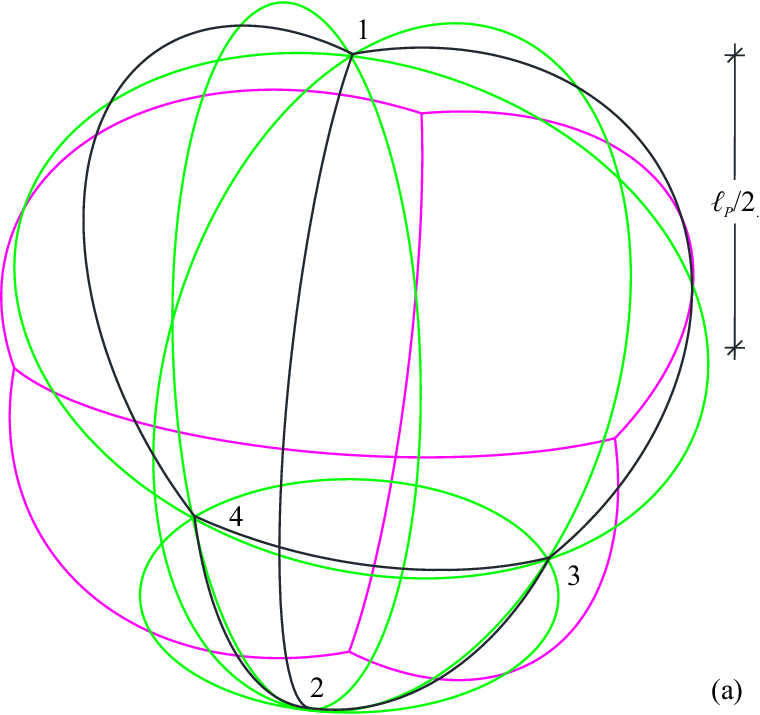}\\
\includegraphics[width=0.4\textwidth]{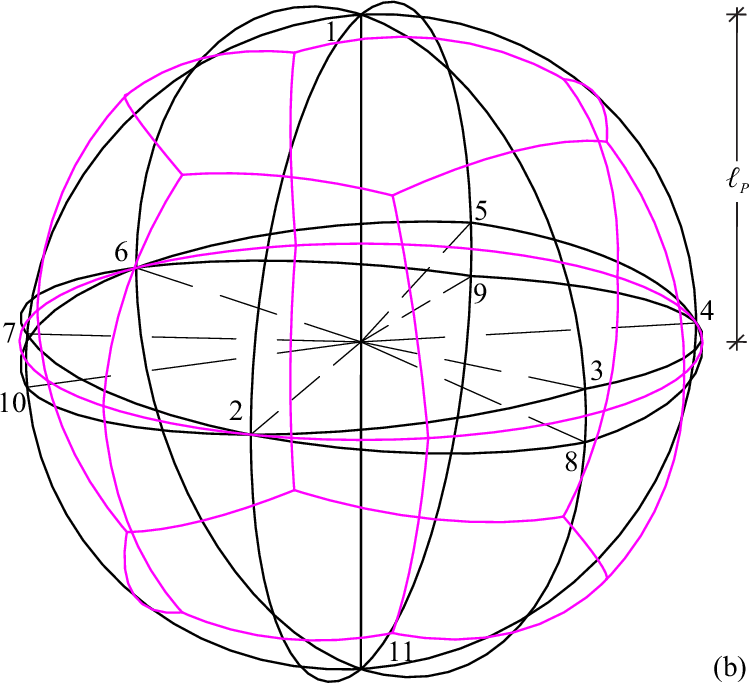}
\caption{\label{Fig_03} Axionometric views of micro-BHs with Voronoi tesselation (pink). (a) Unipolar $\pi$-bit BH, and (b) bipolar $4\pi$-bit BH.}
\end{figure}

However, the definition \eqref{13} is inappropriate for small BHs. Two micro-BHs in $\mathbb{R}^3$ with triangulated event horizons are shown in Fig.~\ref{Fig_03}, wherein $\pi$-bit BH ($R_{BH}=\ell_P/2)$ comprises four vertices and $4 \pi$-bit BH $\left(R_{BH}=\ell_P\right)$ comprises eleven vertices with the fractional part \eqref{52} containing six spherical triangles around the equator. As shown, only the latter event horizon defines only one, precise diameter relation between its two poles 1 and 11. Notably, neither a 1-bit nor a 2-bit BH allows triangulation.

The Euclidean space $\mathbb{R}^{n}$, researched as a simplicial $n$ manifold (i.e., triangulated for $n=2$ ), inherits a natural topology from $\mathbb{R}^{n}$. Moreover, such an approach disentangles the topological (metric-independent) and geometric (metric-dependent) content of the modeled quantities \cite{16, 17}. Therefore, Planck areas on holographic spheres and BHs horizons must be spherical triangles. This is compatible with the Causal Dynamical Triangulation (CDT) approach, which does not assume a pre-existence of a dimensional space, but focuses on the evolution of the spacetime as such.

BHs are by no means stable entities, as discussed in the subsequent section.

\section{Black Hole Diameter Fluctuations}\label{sec:4}
BH energy increases if a photon of energy $E=h c / \lambda=M c^{2}$ is incident on a BH event horizon. If we intend to lose all photon information, it should have a sufficiently long wavelength to ensure that the point of collision with the horizon is uncertain \cite{18}, \cite{19} (p. 160). Therefore, its wavelength must be greater than or equal to the BH radius \eqref{26}. This thought experiment was famously devised by Jacob Bekenstein to derive the BH entropy \eqref{44}. If we assume the condition of equality $\left(\lambda=R_{BH}\right)$, the Compton mass of the photon is $M=h /\left(c R_{BH}\right)$, and hence, the new BH radius would be
\begin{equation}\label{14}
R_{BH}^{A}=R_{BH}+\delta R=\frac{2 G M_{BH}}{c^{2}}+\frac{2 G}{c^{2}} \frac{h}{c R_{BH}},
\end{equation}
the new BH area would be
\begin{equation}\label{15}
A_{BH}^{A}=A_{BH}+\delta A=4 \pi\left(R_{BH}+\delta R\right)^{2},
\end{equation}
and the new BH information capacity would be
\begin{equation}\label{16}
N_{BH}^{A}=64 \pi^{3} \frac{\ell_{p}^{2}}{R_{BH}^{2}}+32 \pi^{2}+4 \pi \frac{R_{BH}^{2}}{\ell_{p}^{2}} .
\end{equation}

\begin{figure}
\includegraphics[width=0.5\textwidth]{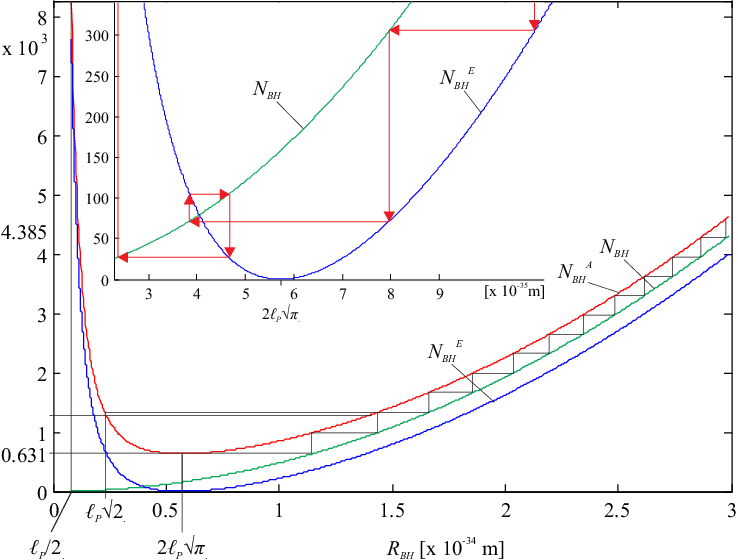}
\caption{\label{Fig_04} Black hole information capacity: $N_{BH}^A$ (red) -- after informationless absorption, $N_{BH}^E$ (blue) -- after informationless emission, $N_{BH}$ (green) -- initial information capacity.}
\end{figure}

Equation \eqref{16} attains the minimum for $R_{BH}=2 \ell_{P} \sqrt{\pi}$, as shown in Fig.~\ref{Fig_04}. The BH with this radius (and mass equal to $m_{P} \sqrt{\pi}$ ) will increase its information capacity fourfold from $N_{BH}=(4 \pi)^{2}$ to $N_{BH}^{A}=(8 \pi)^{2}$ after absorbing the photon with mass $M=m_{P} \sqrt{\pi}$. Furthermore, transmission of another photon with a wavelength corresponding to the new BH radius will increase its information capacity. For radii distinct from $2 \ell_{P} \sqrt{\pi}$, the same BH information capacity can be obtained for two distinct initial radii, and consequently, for two distinguished photon energies.

The informationless BH absorption \eqref{16} can be reversed to produce the informationless BH emission: we do not intend to know the point of emission on the event horizon. Therefore, by subtracting $\delta R$ in Equations \eqref{14}-\eqref{16} we get
\begin{equation}\label{17}
N_{BH}^{E}=64 \pi^{3} \frac{\ell_{p}^{2}}{R_{BH}^{2}}-32 \pi^{2}+4 \pi \frac{R_{BH}^{2}}{\ell_{p}^{2}},
\end{equation}
which defines the BH information capacity after emitting a photon with a wavelength corresponding to its radius. The emission was explained by Hawking \cite{20} and contributes to collapsibility of micro-BHs.

The informationless BH emission \eqref{17} fluctuates around $R_{BH}=2 \ell_{P} \sqrt{\pi}$, shown in the inset of Fig.~\ref{Fig_04}, and increases for $R_{BH}<2 \ell_{P} \sqrt{\pi}$, despite the emission of the wavelength.

These fluctuations should be considered along with the BH informationless absorption \eqref{16}. Upon considering both these processes together, the BH mass along with its diameter fluctuates with time.

\begin{figure}[h]
\includegraphics[width=0.5\textwidth]{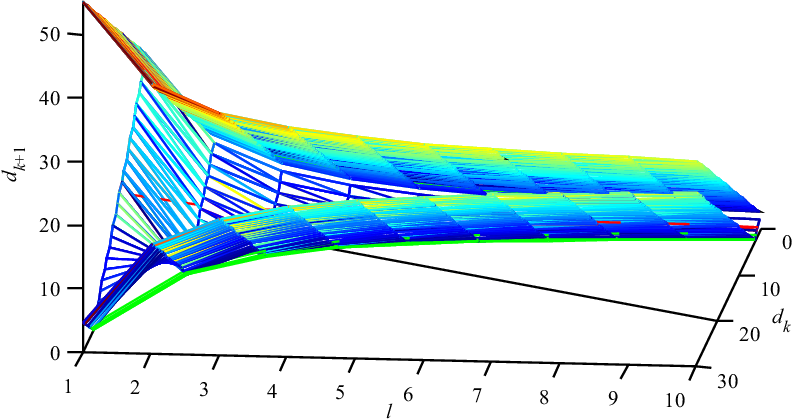}
\caption{\label{Fig_05} New $2$-sphere BH diameter $d_{k+1}$ having initial diameter $d_k$ after absorption or emission of a wavelength $l$. The sinusoidal characteristic of collapse is due to $l$ drawn as a natural number. For the informationless case, shown in Fig.~\ref{Fig_04}, $d = 2l$.}
\end{figure}

BH informationless absorption \eqref{16} and emission \eqref{17} can be extended (cf. Appendix \ref{app:3}) to arbitrary wavelengths $l$ (in general, shedding the informationless property) as
\begin{equation}\label{18}
N_{BH}^{A / E}(d, l)=64 \pi^{3} \frac{1}{l^{2}} \pm 16 \pi^{2} \frac{d}{l}+\pi d^{2},
\end{equation}
which can be described more compactly $\left(N_{BH}=\pi d_{BH}^{2}\right)$ as a recurrence relation
\begin{equation}\label{19}
d_{k+1}^{A/E}=\sqrt{64 \pi^{2} \frac{1}{l^{2}} \pm 16 \pi \frac{d_{k}}{l}+d_{k}^{2}},
\end{equation}
where $d_{k+1}$ is a new diameter of the BH (2-sphere) with an initial diameter $d_{k}$ after absorption ($A$, ''$+$'') or emission ($E$, ''$-$'') of a mass with the Compton wavelength $l$, as shown in Fig.~\ref{Fig_05}.

This derivation can be extended to other dimensions $n$, thereby yielding the following recurrence relation for a BH, $(n-1)$-sphere, diameter after absorption ($A$, ''$+$'') or emission ($E$, ''$-$'') of a wavelength $l$
\begin{equation}\label{20}
\left(d_{k+1}^{A/E}\right)^{n-1}=\left(d_{k} \pm \frac{8 \pi}{l}\right)^{n-1} .
\end{equation}
The RHS of Equation \eqref{20} shows an initial state, whereas the LHS shows the final state of a BH after absorption/emission of a wavelength $l$.

\begin{table}[h]
\caption{\label{table2} Black Hole information capacity $N_{BH}$ in integer dimensions $n=-7,-5,\ldots,9$.}
\begin{ruledtabular}
\begin{tabular}{|l|l|l|l|}
$n$ & $N_{BH}(n)$ & $n$ & $N_{BH}(n)$ \\
\hline
$-7$ & $3360/(\pi^4 d^8)$ & $3$ & $\pi d^2$ \\
$-5$ & $-240/(\pi^3 d^6)$ & $4$ & $\pi^2 d^3/4$ \\
$-3$ & $-24/(\pi^2 d^4)$ & $5$ & $\pi^2 d^4/6$ \\
$-1$ & $-4/(\pi d^2)$ & $6$ & $\pi^3 d^5/32$ \\
$0$ & $0$ & $7$ & $\pi^3 d^6/60$ \\
$1$ & $2$ & $8$ & $\pi^4 d^7/384$ \\
$2$ & $\pi d$ & $9$ & $\pi^4 d^8/840$
\end{tabular}
\end{ruledtabular}
\end{table}

Following the limits of one-sided thermodynamic equilibrium, the BHs can certainly be studied in ($n + 0i$)-dimensions with $n\neq3$, which renders them as markers of dimensionality. Table~\ref{table2} summarizes the information capacities as the functions of the diameter multiplier $d$ for BHs in a range of integer dimensions \cite{21}. We shall return to it in Section \ref{sec:7}. Particularly
\begin{equation}\label{21}
N_{BH}(n) N_{BH}(2-n)=4 \sin \left( \frac{\pi n}{2} \right) .
\end{equation}
There are at least three distinct thresholds for a BH collapse after the emission of wavelength $l$, which can be postulated in Equation \eqref{20}:
\begin{enumerate}
\item $d_{k+1}^{E}=0$ --- total collapse,
\item $d_{k+1}^{E}=1 /(2 \pi)$ --- the Planck temperature collapse, and
\item $d_{k+1}^{E}=1 / \sqrt{\pi}$ --- one-bit collapse.
\end{enumerate}
The total BH collapse occurs after the emission of wavelength $l=8 \pi / d_{k}$, i.e., after the emission of its entire Compton mass \eqref{9}. For Sagittarius A$^*$, $l \leq \num{1.66E-44} \ll 1$, so Sagittarius A$^*$ will never collapse.

If we set $l \geq 1$ in Equation \eqref{20}, we obtain
\begin{equation}\label{22}
d_{k} \le 8 \pi \approx 25.132,
\end{equation}
for $d_{k+1}^{E}=0$,
\begin{equation}\label{23}
d_{k} \le 8 \pi+\frac{1}{2 \pi} \approx 25.292,
\end{equation}
for $d_{k+1}^{E}=1 / 2 \pi$, and
\begin{equation}\label{24}
d_{k} \le 8 \pi+\frac{1}{\sqrt{\pi}} \approx 25.697,
\end{equation}
for $d_{k+1}^{E}=1 / \sqrt{\pi}$, which are the thresholds for the maximum diameter multipliers of collapsible BHs. The lowest threshold $d_{k}=8 \pi$ corresponds to a BH with mass $2 \pi$ times greater than the Planck mass. This threshold has already been derived as BH threshold of distinguishability \eqref{10} by considering the BH Compton wavelength and shall be further delivered in Equation \eqref{89} by considering the variational potential \eqref{78} on the event horizon \eqref{88}.

\begin{table}[ht]
\caption{\label{table3} Emission wavelength $l$ maintaining constant BH diameter as in integer dimensions $n = –7, –6,\ldots,9$.}
\begin{ruledtabular}
\begin{tabular}{|l|l|}
$n$ & Emitted $l$ maintaining constant BH diameter \\
\hline
$-7, 9$ & $\begin{aligned}
&l_1=\frac{4 \pi}{d_k}, l_{2,3,4,5}=\frac{4 \pi}{d_k}(1 \pm i \sqrt{3 \pm 2 \sqrt{2}}) \\
&l_{6,7}=\frac{4 \pi}{d_k}(1 \pm i)
\end{aligned}$ \\
$-6, 8$ & complicated and complex solution \\
$-5, 7$ & $l_1=\frac{4 \pi}{d_k}, l_{2,3}=\frac{4 \pi}{d_k}(1 \pm \sqrt{3} i), l_{4,5}=\frac{4 \pi}{d_k}\left(1 \pm \frac{\sqrt{3}}{3} i\right)$ \\
$-4, 6$ & $l_{1,2,3,4}=\frac{4 \pi}{d_k}\left(1 \pm \frac{\sqrt{5 \pm 2 \sqrt{5}}}{\sqrt{5}} i\right)$ \\
$-3, 5$ & $l_1=\frac{4 \pi}{d_k}, l_{2,3}=\frac{4 \pi}{d_k}(1 \pm i)$ \\
$-2, 4$ & $l_{1,2}=\frac{4 \pi}{d_k}(1 \pm \frac{\sqrt{3}}{3} i)$ \\
$-1, 3$ & $l_1=\frac{4 \pi}{d_k}$ \\
$0, 2$ & contradiction ($1\neq0$) \\
$1$ & identity ($1=1$)
\end{tabular}
\end{ruledtabular}
\end{table}

The BH diameter does not vary ($d_{k+1}^{E}=d_{k}$) after the emission of wavelengths listed in Table~\ref{table3}. Equation \eqref{20} is symmetrical  with respect to $n=1$, i.e., the wavelengths solving $d_{k+1}^{E}=d_{k}$ are the same for $n$ and $2-n$, which is a reflection relation around 2. According to prior research \cite{22}, an ordinary space of $(m-2)$-dimension is equivalent to the $m$-dimensional superspace with an anticommuting coordinate, and a 4-space with only one anticommuting coordinate (time) is equivalent to an ordinary space with negative dimensions of $-2$.

The analytical formula $l(m)$ for constant diameter emission in any dimension, if one exists, remains to be researched. The odd dimensionalities admit the real wavelength $l=4 \pi / d_{k}$ corresponding to precisely half the BH Compton wavelength \eqref{9}. In even dimensionalities, all constant diameter wavelengths $l$ are complex, thereby indicating the existence of a link with vanishing real volumes and surfaces of $m$-balls, in even, negative dimensions (owing to the vanishing rational factor $g_{m}$, not because of the vanishing diameter) \cite{21}.

\section{Entropic work of gravity}\label{sec:5}
Entropic gravity defines that both inertia and gravity are phenomena emergent on a holographic screen, whereas the classical concepts such as position, velocity, acceleration, mass, and force\footnote{Time is not included in this enumeration (\cite{23}, p. 2).} are far from obvious \cite{23}. Quoting Erik Verlinde, ''when a particle has an entropic reason [$\delta S$] to be on one side [displaced by $\delta R$] of the membrane [holographic screen] and the membrane carries a temperature [$T$], it [the particle] will experience an effective force [$F$] equal to $[\delta W=] F \delta R=T \delta S$'' \cite{23} (Equation 3.3).

Based on this entropic work formula, the Second law of Newton \cite{23} (Equation 3.5) can be recovered by combining the Unruh temperature $T$ \cite{23} (Equation 3.4) with the postulated variation of entropy $\delta S$ near the screen \cite{23} (Equation 3.2), linearized with respect to the reduced Compton wavelength $\lambdabar=\hbar / (M c)$ by introducing inertial mass $M$ to this equation. Newton's law of gravity \cite{23} (Equation 3.9) can be recovered by assuming a sphere partitioned into Planck areas $\ell_{P}^{2}$. Instead of the Unruh effect, the temperature is derived from the equipartition theorem \cite{23} (Equation 3.7) for a degree of freedom, multiplied by the number of these areas. Particularly, the mass-energy equivalence $E=M_{2} c^{2}$ \cite{23} (Equation 3.8) introduces the $2^{\text {nd }}$ gravitational mass $M_{2}$ to the equation. Entropic, emergent gravity \cite{23} has been experimentally confirmed to date \cite{24}.

The same entropic work equation can be derived \cite{14} using an entropy formula \cite{14} (Equation 1) that reduces to the Bekenstein–Hawking entropy at the event horizon. As such, the temperature \cite{14} (Equation 2) is obtained using the Hawking formula and expressed as the gradient of the gravitational potential of the $2^{\text {nd }}$ gravitational mass $M_2$, which further reduces to the BH temperature at the event horizon. This derivation is certainly compelling to those agreeing with Verlinde’s perspective concerning the entropic origin of gravity and inertia and provides new insights into this matter. Let us delve more closely into this derivation.

\begin{figure}[h]
\includegraphics[width=0.5\textwidth]{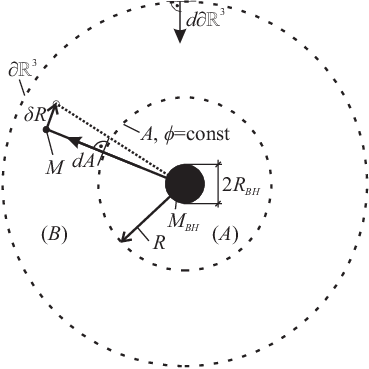}
\caption{\label{Fig_06} Simplified model of the universe used in this study.}
\end{figure}

As shown in Fig.~\ref{Fig_06}, we considered a model of a \emph{closed} 3-dimensional spherical, universe $\mathbb{R}^{3}$ with an edge of $\partial \mathbb{R}^{3}$. The concept of an edge $\partial \mathbb{R}^{3}$ is certainly a strong abuse of notation and does not indicate that the Euclidean space $\mathbb{R}^{3}$ is closed. However, this concept can be introduced in Equation \eqref{36}, for instance, to prove that the surface integral vanishes if the edge $\partial \mathbb{R}^{3}$ is moved to infinity. Regardless, we cannot study Euclidean space $\mathbb{R}^{3}$ without a universe from which to study it. Based on our current knowledge, the radius of the observable universe is approximately 46 billion light years.

At its center, our toy universe contains a Schwarzschild BH with (passive) mass $M_{BH}$. Its presence is much more formally unsettling than that of the edge $\partial \mathbb{R}^{3}$, considering it does not contain an interior.

Passive mass $M_{BH}$ generates a conservative and irrotational gravitational field that can be expressed in terms of a scalar, gravitational potential $\phi_{g}$ as
\begin{equation}\label{25}
\phi_{g}=-\frac{G M_{BH}}{R},
\end{equation}
for $R \geq R_{BH}$, where
\begin{equation}\label{26}
R_{BH}=\frac{2 G M_{BH}}{c^{2}},
\end{equation}
represents the radius of the BH event horizon.

The minimum potential $\phi_{g}$ occurs at the event horizon and amounts to
\begin{equation}\label{27}
\phi_{BH}=-\frac{G M_{BH}}{R_{BH}}=-G M_{BH} \frac{c^{2}}{2 G M_{BH}}=-\frac{1}{2} c^{2} .
\end{equation}
An equipotential surface $A$ surrounds the $M_{BH}$ in a volume $(A)$ for $R \geq R_{BH}$. Thus, the single-mass $M_{BH}$ system is described by Laplace's equation
\begin{equation}\label{28}
\nabla^{2} \phi_{g}=0 .
\end{equation}
Let us now consider a test mass $M$ located in the background of $M_{BH}$ in volume $(B)$, such that $(A)+(B)=\mathbb{R}^{3}$ and $B=A \cup \partial \mathbb{R}^{3}$. Owing to this simplified model of a spherical universe including just two masses, we do not encounter the three-body problem, and yet this model can adequately demonstrate all the current findings of this study. Moreover, owing to its spherical symmetry, $R$ is the only significant dimension. The system of these two masses $M_{BH}$ and $M$ can be described using Poisson's equation for gravity
\begin{equation}\label{29}
\nabla^{2} \phi_{g}=4 \pi G \rho,
\end{equation}
where $\rho$ is the mass density in the volume $(B)$ comprising $M$. Poisson's equation \eqref{29} is the inhomogeneous version of Laplace's equation \eqref{28} and includes the latter if $\rho=0$. It is satisfied on any equipotential surface $A$, and integrating over $(A)$ using the divergence theorem yields
\begin{equation}\label{30}
4 \pi G \iiint_{(A)} \rho d V_{A}=\iiint_{(A)} \nabla^{2} \phi_{g} d V_{A}=\varoiint_{A} \nabla_{R} \phi_{g} d A,
\end{equation}
where $\nabla_{R}$ is the gradient of the gravitational potential \eqref{25}, which is the derivative in the direction of the normal vector on the surface $A$ pointing outside the volume $(A)$.

Newton's law of gravity \eqref{48} is a general physical law derived from empirical observations. In this case, the passive mass $M_{BH}$ is located at the center of the simplified model of the spherical universe, which implies that the spherically symmetric gravitational potential $\phi_{g}(R)$ \eqref{25} and mass density $\rho(R)$ satisfy Laplace's equation \eqref{28}. The $2^{\text {nd}}$ mass $M$ and all the subsequent test masses introduced to this model would only induce mass density inhomogeneity and attain the same equilibrium in a \emph{free fall} process induced by $\nabla_{R} \phi_{g}$, unless some work is done to disturb it.

Such a disturbing force $F$ - applied to wiggle the test mass $M$ location by a certain $\delta R$ in a time period $\delta t$ (doing work $\delta W$ - equals to the corresponding variation of the potential energy in volume $(B)$, expressed as
\begin{equation}\label{31}
\begin{split}
\delta W &=F \circ \delta R=\delta U=-\iiint_{(B)} \phi_{g} \delta \rho d V_{B}=\\
&=-\frac{1}{4 \pi G} \iiint_{(B)} \phi_{g} \nabla^{2} \delta \varphi d V_{B},
\end{split}
\end{equation}
where  Poisson's equation \eqref{29} is used to express the variation in density $\delta \rho$ in the volume $(B)$ containing, the - now active - mass $M$ with the corresponding variation $\delta \varphi$ of the potential. Let us refer to $\delta R$ as the disturbing radius of the active mass $M$.

Accordingly, integrating variation of the potential energy \eqref{31} over $(B)=\mathbb{R}^{3}-(A)$, we get
\begin{equation}\label{32}
-4 \pi G \delta U=\iiint_{\mathbb{R}^{3}} \phi_{g} \nabla^{2} \delta \varphi d V-\iiint_{(A)} \phi_{g} \nabla^{2} \delta \varphi d V_{A},
\end{equation}
and introducing the $2^{\text {nd }}$ and $3^{\text {rd }}$ terms that negate each other, we get
\begin{equation}\label{33}
\begin{aligned}
-4 \pi G \delta U &=\iiint_{\mathbb{R}^{3}} \phi_{g} \nabla^{2} \delta \varphi d V-\iiint_{(A)} \delta \varphi \nabla^{2} \phi_{g} d V_{A}+\\
&+\iiint_{(A)} \delta \varphi \nabla^{2} \phi_{g} d V_{A}-\iiint_{(A)} \phi_{g} \nabla^{2} \delta \varphi d V_{A}.
\end{aligned}
\end{equation}
Thereafter, we use Green's second identity to transform the $3^{\text {rd }}$ and $4^{\text {th }}$ terms into surface integrals, as
\begin{equation}\label{34}
\begin{aligned}
-4 \pi G \delta U=& \iiint_{\mathbb{R}^{3}} \phi_{g} \nabla^{2} \delta \varphi d V-\iiint_{A} \delta \varphi \nabla^{2} \phi_{g} d V_{A}+\\
&+\varoiint_{A}\left(\delta \varphi \nabla_{R} \phi_{g}-\phi_{g} \nabla_{R} \delta \varphi\right) d A.
\end{aligned}
\end{equation}
We integrate the $4^{\text{th}}$ term over the equipotential surface $A$ to displace the potential $\phi_{g}$ before the integral, and apply the divergence theorem to retrace to the volume integration over $(A)$. However, the $4^{\text{th}}$ term vanishes upon employing Poisson's equation \eqref{29}, considering $\delta \varphi$ contains no sources inside $(A)$. Subsequently, we re-express the $2^{\mathrm{nd}}$ term with the volume integration over $(A)$ as an integration over $(A)=\mathbb{R}^{3}-(B)$
\begin{equation}\label{35}
\begin{aligned}
-4 \pi G \delta U &=\iiint_{\mathbb{R}^{3}} \phi_{g} \nabla^{2} \delta \varphi d V-\iiint_{\mathbb{R}^{3}} \delta \varphi \nabla^{2} \phi_{g} d V+\\
&+\iiint_{(B)} \delta \varphi \nabla^{2} \phi_{g} d V_{B}+\varoiint_{A} \delta \varphi \nabla_{R} \phi_{g} d A.
\end{aligned}
\end{equation}
Similarly, the $3^{\text {rd }}$ term vanishes considering $\phi_{g}$ includes no sources inside $(B)$. Therefore, using Green's second identity to transform the volume integral over $\mathbb{R}^{3}$ into the surface integral over $\partial \mathbb{R}^{3}$, we get
\begin{equation}\label{36}
\begin{aligned}
-4 \pi G \delta U &=\varoiint_{\mathbb{R}^{3}}\left(\phi_{g} \nabla_{R} \delta \varphi-\delta \varphi \nabla_{R} \phi_{g}\right) d \partial \mathbb{R}^{3}+\\
&+\varoiint_{A} \delta \varphi \nabla_{R} \phi_{g} d A,
\end{aligned}
\end{equation}
where the $1^{\text {st }}$ surface integral over $\partial R^{3}$ vanishes as well. Therefore,
\begin{equation}\label{37}
F \circ \delta R=\delta U=-\frac{1}{4 \pi G} \varoiint_{A} \nabla_{R} \phi_{g} \delta \varphi d A.
\end{equation}

This innovative procedure conveniently transformed the volume integration \eqref{31} over $(B)$ defining the universe $\mathbb{R}^{3}$ with a hole $(A)$ into a surface integration \eqref{37} over the equipotential surface $A$ describing only this hole, which expresses the BH event horizon for a limiting case of $R=R_{BH}$.

Both the uniform inertial, that is flat spacetime acceleration $(a)$ and the gravitational acceleration $(g)$ can be related to temperature $T$ according to Unruh blackbody-radiation equation
\begin{equation}\label{38}
T_{a}=\frac{\hbar}{2 \pi c k_{B}} a,
\end{equation}
or the analogous Hawking blackbody-radiation equation
\begin{equation}\label{39}
T_{g}=\frac{\hbar}{2 \pi c k_{B}} g .
\end{equation}
Thus, the temperature can be expressed in terms of the gravitational potential gradient \eqref{25} as
\begin{equation}\label{40}
T_{g}=\frac{\hbar}{2 \pi c k_{B}} \nabla_{R} \phi_{g} .
\end{equation}
To eventually derive Verlinde's entropic gravity, the variation of entropy on $A$ \cite{14} (Equation 1)\footnote{ The additive constant in this equation can represent the fractional part $\{N_A\}$ \eqref{52} of an equipotential surface information capacity.} should be expressed using another function
\begin{equation}\label{41}
\delta S=-\frac{c k_{B}}{2 G \hbar} \delta \varphi A,
\end{equation}
which expresses the entropy variation as a function of the variation of the potential $\delta \varphi$ induced upon relocating the active mass $M$ in volume $(B)$ by $\delta R$, which corresponds to the variation in density $\delta \rho$ in volume $(B)$, according to the variation of the potential energy \eqref{31}
\begin{equation}\label{42}
\delta d S=-\frac{c k_{B}}{2 G \hbar} \delta \varphi d A ,
\end{equation}
on the surface element $d A$ of the surface $A$.
Thereafter, $\nabla_{R} \phi_{g}$ from Equation \eqref{40} and $\delta \varphi$ from Equation \eqref{42} were substituted into Equation \eqref{37}, which demonstrates that this is equal to the entropic work
\begin{equation}\label{43}
\begin{aligned}
F \circ \delta R &=-\frac{1}{4 \pi G} \varoiint_{A}\left(\frac{2 \pi c k_{B}}{\hbar} T_{g}\right)\left(-\frac{2 G \hbar}{c k_{B}} \delta d S\right)=\\
&=\varoiint_{A} T_{g} \delta d S=T_{g} \varoiint_{A} \delta d S=T_{g} \cdot \delta S,
\end{aligned}
\end{equation}
where $T_{g}$ can be displaced before the integral, due to integrating over an equipotential surface $A$. Therefore, work $\delta W$ and force $F$ can be deemed entropic. The author conjectures that all forces are entropic. Thus, work is both the scalar product of force and variation of location and the regular product of temperature and variation of entropy. However, this does not preclude additional variations of force and temperature.

As discussed in Section \ref{sec:7}, the concept of binary potential \eqref{54} transforms entropy variation \eqref{41} on the equipotential surface $A$ into a binary entropy variation \eqref{55} on the holographic screen $A$, generalizing the entropy variation \eqref{41}. The author of \cite{14} reported that the ''number $N$ of 'bits' on the screen made no appearance here'' (i.e., in the derivation of entropic work \eqref{43}), apparently disregarding the importance of a ''bit'' in relation to the BH information paradox. Entropy variation \eqref{41} certainly does not relate to electrodynamics \cite{14}, as it reduces to BH entropy \cite{18} (using \eqref{27}) as
\begin{equation}\label{44}
S_{BH}=-\frac{c k_{B}}{2 G \hbar} \phi_{BH} A_{BH}=\frac{k_{B} c}{2 G \hbar} \frac{c^{2}}{2} 4 \pi R_{BH}^2=\frac{1}{4} k_{B} N_{BH},
\end{equation}
which is a thermodynamical term. Notably, the entropy of any BH is equal to that of the Schwarzschild BH with the same area $A_{BH}$ (by the Penrose process).

BH entropy \eqref{44} can be extended to entropy variation \eqref{41} beyond the event horizon for an arbitrary equipotential $A$ and $\delta \varphi$. However, for radii smaller than the Schwarzschild radius \eqref{26}, it is meaningless and hence disproves assumption $\mathbf{B}$ of \cite{14} that $\mathbb{R}^{3}$ is made up of the union of nonintersecting holographic screens. Screens \emph{passing across} BHs cannot be defined, as they would have created voids in such a \emph{holographized} $\mathbb{R}^{3}$. Thus, if we relied only on the intuitive concept of \emph{space}, we could use neither the divergence theorem nor Green's second identity in the derivation entropic work \eqref{43}. Of course, these mathematical theorems are valid, and we have used them correctly. Only our intuition is misleading.

Neither Verlinde's \eqref{46} nor Hossenfelder's \eqref{41} entropy variations should be considered as an entropic explanation of any physically observed inverse-square law. Both these equations are related, which will be discussed in the subsequent section.

\section{Entropy Variation Sphere and Variational Potential}\label{sec:6}
We compare Hossenfelder's entropy variation \eqref{41} with Verlinde's entropy variation in the vicinity of a holographic screen \cite{23} (Equation 3.1)
\begin{equation}\label{45}
\delta S=2 \pi k_{B} \quad \text { when } \quad \delta R=\frac{\hbar}{Mc},
\end{equation}
or \cite{23} (Equation 3.2)
\begin{equation}\label{46}
\delta S=2 \pi k_{B} \frac{Mc}{\hbar} \delta R=2 \pi k_{B} \frac{\delta R}{\lambdabar_{M}},
\end{equation}
in the case of a test mass $M$ approaching the screen. The factor of $2 \pi$ included in Equations \eqref{45}, \eqref{46} is apparently related to the \emph{threshold of distinguishability} \eqref{6}.

Entropy variation \eqref{46} enables the recovery of the second law of Newton, plugging $a$ from the Unruh temperature \eqref{38}
\begin{equation}\label{47}
\begin{aligned}
F \circ \delta R &=T_{a} \cdot \delta S=\frac{\hbar}{2 \pi c k_{B}} a \cdot 2 \pi k_{B} \frac{Mc}{\hbar} \delta R=\\
&=Ma \cdot \delta R \Leftrightarrow F=Ma=M \frac{\delta R}{\delta t^{2}},
\end{aligned}
\end{equation}
and Newton's law of gravity, by expressing Hawking temperature \eqref{39} in terms of the gradient of gravitational potential \eqref{25} of the passive mass $M_{BH}$
\begin{equation}\label{48}
\begin{aligned}
F \circ \delta R&=T_{g} \cdot \delta S=\frac{\hbar}{2 \pi c k_{B}}\left(\frac{G M_{BH}}{R^{2}}\right) \cdot 2 \pi k_{B} \frac{Mc}{\hbar} \delta R=\\
&=\frac{G M_{BH} M}{R^{2}} \cdot \delta R \Leftrightarrow F=\frac{G M_{BH} M}{R^{2}}.
\end{aligned}
\end{equation}
Alternatively, the original Verlinde's procedure can be used with Hawking temperature \eqref{39} induced by the equipartition theorem for a single degree of freedom \eqref{166} and mass-energy equivalence \eqref{64}, as described at the beginning of this Section. Furthermore, Verlinde's entropy variation \eqref{46} directly introduces the Unruh temperature \eqref{38} as
\begin{equation}\label{49}
\begin{aligned}
&M a \circ \delta R=T_{a} \cdot \delta S=T_{a} \cdot 2 \pi k_{B} \frac{M c}{\hbar} \delta R \Leftrightarrow \\
&a=\frac{2 \pi c k_{B}}{\hbar} T_{a}.
\end{aligned}
\end{equation}
However, none of these laws can be recovered from Hossenfelder's entropy variation \eqref{41}, the starting point of which is gravitational potential \eqref{25} analyzed in the context of Poisson's equation for gravity \eqref{29}. Therefore, the current approach investigated the recovery of these laws.

Let us define the potential variation on the equipotential surface $A$, induced by wiggling the active mass $M$ location by $\delta R$, as gravitational potential \eqref{25} (with ''$-$'')
\begin{equation}\label{50}
\delta \varphi \doteq-\frac{G M}{\delta R}=-\frac{G}{\delta R} \frac{\hbar}{c \lambdabar_{M}} .
\end{equation}
This ensures that the equipotential surface $A$ is a sphere of the disturbing radius $\delta R$. We shall further refer to this potential as the \emph{variational potential} and the corresponding sphere as the \emph{entropy variation sphere}. By substituting variational potential \eqref{50} into entropy variation \eqref{41}, we recover Verlinde's entropy variation \eqref{46} as
\begin{equation}\label{51}
\delta S=-\frac{c k_{B}}{2 G \hbar}\left(-\frac{G M}{\delta R}\right) 4 \pi \delta R^{2}=2 \pi k_{B} \frac{M c}{\hbar} \delta R,
\end{equation}
and hence, we recover also Newton's second law of motion \eqref{47} and Newton's law of gravity \eqref{48}.

The hypothesis concerning the variational potential \eqref{50} is a consequence of the entropic work derivation presented in \cite{14}. At the onset, $A$ is assumed to surround $M_{BH}$ in $\mathbb{R}^{3}$, as shown in Fig.~\ref{Fig_06}, and is not related to $\delta R$. However, after completing the derivation, $A$ represents solely a 2-sphere, a set of points bound by a 2-sphere diameter definition \eqref{13} situated at a fixed distance from a BH event horizon. Entropy variation \eqref{41} is the function of the potential variation on the equipotential surface $A$, whereas entropy variation \eqref{46} was motivated by Bekenstein's original thought experiment \cite{18}, which resulted in the BH entropy \eqref{44}. As such, they describe the equipotential sphere from its opposite sides, which is reflected by the inversion of their signs. The side described by entropy variation \eqref{41} is the side of the passive mass $M_{BH}$, wherein the side described by entropy variation \eqref{46} represents the side of the active mass $M$, which can be observed by deriving BH entropy \eqref{44} from Verlinde's entropy variation \eqref{46} (cf. Appendix \ref{app:6}). These sides are related \eqref{51} through the entropy variation sphere, which introduces the sign reversal. In such a radial configuration, a quantity is positive if it is consistent with the radius; else, it is negative. In particular, $M_{BH}$, $M$, $R$, and $\delta R$ are parameters of this holographic 2-sphere. Moreover, $R$ denotes a time-independent version of $\delta R$, as discussed in the subsequent section.

\section{Binary Potential and Binary Messages}\label{sec:7}
Entropy is a measure of information (even if it is missing information \cite{25} or a measure of surprise \cite{vedral}) and information is measured in bits. According to the holographic principle, the bounded interior and unbounded exterior are separated by a two-dimensional boundary, and by definition, each $k^{\text {th }}$ bit is physically represented on this boundary (the holographic sphere $A$) by the Planck area \cite{13,18,23}. For area $A$, the boundary exhibits the information capacity of $N_{A}=A / \ell_{P}^{2}$ bits. However, this is an imprecise conclusion, considering $N_{A} \in \mathbb{R}$ and the number of bits is a natural number. Moreover, for a spherical surface, $A=4 \pi R^{2}$, $N_{A}$ is a transcendental number: a quotient of the product of a transcendental number $(\pi)$ with a rational number $\left(4 R^{2}\right)$ and a rational number $\left(\ell_{P}^{2}\right)$.

Therefore, the number of bits provided by the boundary is $\left\lfloor N_{A}\right\rfloor \in \mathbb{N}_{0}$, where ''$\left\lfloor \right\rfloor$'' is the floor function, that yields the greatest integer less than or equal to $N_{A}$. Thus, the fractional part
\begin{equation}\label{52}
0<\left\{N_{A}\right\}<1,
\end{equation}
can be considered as a signature of the holographic sphere $A$. Perhaps the continuum hypothesis ensures a unique $\left\{N_{A}\right\}$ for any given holographic sphere, regardless of the simultaneous existence of the same number of bits $N_{A}$ on an infinitely countable number of other spheres.

Thus, entropy variation \eqref{41} can be expressed in terms of the information capacity of the holographic sphere $A$ as
\begin{equation}\label{53}
\delta S=-\frac{c k_{B}}{2 G \hbar} \delta \varphi N_{A} \ell_{P}^{2}=-\frac{c k_{B}}{2 G \hbar} \delta \varphi N_{A} \frac{\hbar G}{c^3}=-\frac{1}{2} k_{B} N_{A} \frac{\delta \varphi}{c^{2}} .
\end{equation}
The last term of entropy variation \eqref{53} corresponds to $\Delta S$ \cite{23} (Equation 3.12), where the Newton potential keeps track of the depletion of the entropy per bit.

Entropy variation \eqref{53} can be further simplified by postulating the binary variational potential (or simply, \emph{binary potential}) in the Planck time $t_{P}$ over the $k^\text{th}$ Planck area $\ell_{P}^{2}$, defined as
\begin{equation}\label{54}
\delta \varphi_{k} \doteq-\{0,1\} \ell_{P}^{2} / t_{P}^{2}=-\{0,1\} c^{2} .
\end{equation}
This relation between the speed of light and the Planck area is remarkable in the context of the recent discovery that the Planck length and the Planck time can be obtained from a Newtonian force-spring knowing only the value of the speed of light and with no knowledge of the gravitational constant $G$ or the Planck constant $h$ \cite{26}. More specifically, the variational potential requires the notion of time, assumed as imaginary, i.e., $(i c)^{2}=-c^{2}$. This unit also stems from the 2-dimensionality of the Planck area. Such a definition of the binary potential must be true if a bit of information is the only property of $\ell_{P}^{2}$ on a holographic screen $A$. Therefore, the equipotential surface $A$ and the holographic screen $A$ are the same 2-spheres.

By substituting binary potential \eqref{54} into entropy variation \eqref{53}, we obtain binary entropy variation on a holographic sphere $A$, given as
\begin{equation}\label{55}
\delta S=-\frac{1}{2} k_{B} N_{A} \frac{\delta \varphi}{c^{2}}=-\frac{1}{2} k_{B} \sum_{k=1}^{\left\lfloor N_{A}\right\rfloor} \frac{\delta \varphi_{k}}{c^{2}}=\frac{1}{2} k_{B} N_{1},
\end{equation}
where the summation covers all Planck areas of $A$, and $N_{1} \in \mathbb{N}$ denotes the number of Planck areas with binary potential $\delta \varphi_{k}=-c^{2}$, further called \emph{active Planck areas}.

We note in passing that only $n = 2$ provides the first possibility of assigning a bit to a circular segment equal to the Planck length. This would mean the lack of black holes in lower integer dimensions. In such a 2-dimensional case the binary potential is conjectured to be $\delta \varphi_{k}(2)=\left\{0,1\right\}ic$, so that its square yields $\delta \varphi_{k}(3)=-\left\{0,1\right\}c^2$ \eqref{54} (on a Planck area spherical triangle). Following this rule $\delta \varphi_{k}(4)=-\left\{0,1\right\}ic^3$ (on a Planck volume spherical tetrahedron), $\delta \varphi_{k}(5)=\left\{0,1\right\}c^4$, $\delta \varphi_{k}(6)=\left\{0,1\right\}ic^5$, etc. Integer dimensional binary potentials are related to each other by integral powers of the imaginary unit:  $\delta \varphi_{k}(n)=\left\{0,1\right\}c^{n-1}i^{n-1}$.

On comparing BH entropy \eqref{44} with the binary entropy variation \eqref{55}, we get
\begin{equation}\label{56}
S_{BH}=\delta S \Leftrightarrow \frac{1}{4} k_{B} N_{BH}=\frac{1}{2} k_{B} N_{1} \Leftrightarrow N_{1}=\frac{1}{2} N_{BH},
\end{equation}
which indicates that the temporary distribution of Planck areas on BH event horizon maximizes Shannon entropy. Therefore, the event horizon is algorithmically random or patternless \cite{27} binary message comprising a balanced number of Planck areas with binary potential \eqref{54} equal to $-c^{2}$ and zero.

An additional, trivial observation is that the average potential of all Planck areas on the event horizon is equal to the event horizon potential at a vertex, i.e.,
\begin{equation}\label{57}
\frac{\sum_{k=1}^{N_1}\left(-c^2\right)+\sum_{k=1}^{\left\lfloor N_{BH}\right\rfloor - N_1} 0}{N_{BH}}=\frac{-N_1 c^2}{N_{BH}}=\frac{-\frac{1}{2} N_{BH} c^2}{N_{BH}}=-\frac{1}{2} c^2.
\end{equation}
The two binary strings
\begin{equation}\label{58}
X_{1}=11111111111111111111,
\end{equation}
\begin{equation}\label{59}
X_{2}=10110100101000101101,
\end{equation}
assumed to be generated by tosses of a fair coin (random variable outcomes) must be considered equally random. However, based on the perspective of Kolmogorov complexity, the second string is more random than the first one considering it is a patternless binary sequence \cite{27}. If strings \eqref{58} and \eqref{59} were messages transmitted bit by bit over a certain transmission channel, the probabilities for zero and one occurrences in the second message \eqref{59} would be equal ($p_{0}=p_{1}=1 / 2$). Therefore the second message would be a noncompressible one, maximizing Shannon entropy, as
\begin{equation}\label{60}
H\left(X_{2}\right)=-\frac{1}{2} \ln \left(\frac{1}{2}\right)-\frac{1}{2} \ln \left(\frac{1}{2}\right)=\ln (2),
\end{equation}
expressed in nats, whereas Shannon entropy of the message \eqref{58} is zero, as it could be compressed to just one bit. The maximal Shannon entropy of the second message \eqref{59} would also minimize the amount of energy required to erase a single bit of information on the event horizon, which is expressed using the Landauer erasure limit \cite{28}
\begin{equation}\label{61}
E_{1}=T_{BH} k_{B} H\left(X_{2}\right)=T_{BH} k_{B} \ln (2).
\end{equation}

On comparing entropy variation \eqref{53} and  binary entropy variation \eqref{55}, the variational potential \eqref{50} can be expressed solely in terms of the information capacity as
\begin{equation}\label{62}
\delta \varphi=-\frac{N_{1}}{N_{A}} c^{2},
\end{equation}
which is an important relation that also (cf. Table~\ref{table5}) shows that the event horizon is a patternless binary message.

Considering the Hawking BH radiation is diameter-dependent black-body radiation bearing no information regarding BH interior conditions, it is a patternless message containing the same number of zeros and $-c^2$ terms ($-c^2/2$ on average), as a binary one-time pad secret key. Holographic screens other than event horizons provide more diversified (and less surprising) distributions of zeros and ones with lower Shannon entropies (further discussed in Section \ref{sec:9}). However, an event horizon is a patternless binary message \eqref{56} and the quantum information of the in-fallen matter cannot be hidden in correlations between the semiclassical Hawking radiation and BH internal states, as asserted by the no-hiding theorem \cite{1}.

The entropic work of a BH event horizon can be expressed using Hawking temperature \eqref{39} and BH entropy \eqref{44}  as
\begin{equation}\label{63}
T_{g} \cdot \delta S=\left(\frac{\hbar}{2 \pi c k_{B}} \frac{G M_{BH}}{R_{BH}^{2}}\right)\left(\frac{k_{B} N_{BH}}{4}\right)=\frac{M_{BH} c^{2}}{2}=\frac{d_{BH} E_{P}}{8},
\end{equation}
which differs from the BH mass-energy equivalence
\begin{equation}\label{64}
E_{BH}=M_{BH} c^{2}=\frac{1}{4} d_{BH} E_{p},
\end{equation}
where $E_{P}$ is the Planck energy, and results in two distinct Landauer bounds on a minimum BH information capacity (cf. Appendix \ref{app:7}): $N_{BH}=2 \ln (2)$ (mass-energy equivalence, 1 bit) and $N_{BH}=4 \ln (2)$ (BH entropic work, 2 bits). To assign one unit of entropy on a BH horizon at least four bits (Planck areas) are required.

Heisenberg's Uncertainty Principle (HUP) is commonly expressed as
\begin{equation}\label{65}
\delta P \delta R \geq \hbar / 2,
\end{equation}
where $\delta P$ is momentum and $\delta R$ is the position uncertainty. In an alternative form, the above equation can be expressed as
\begin{equation}\label{66}
\delta E \delta t \geq \hbar / 2,
\end{equation}
where $\delta E$ is energy and $\delta t$ is the time uncertainty. The energy uncertainty $\delta E$  of one bit can be expressed by the temperature uncertainty using the equipartition theorem ($\delta E = k_{B} \delta T / 2$). Substituting this with the time uncertainty discretized by the Planck time ($\delta t = t~t_P$, $t \in \mathbb{R}$) into Equation \eqref{66} we get the temperature uncertainty
\begin{equation}\label{67}
\delta T \geq \frac{1}{t}T_{P},
\end{equation}
which is equal to the Planck temperature if $t = 1$. Thus, if the time uncertainty $\delta t$ of measurement of one bit equals the Planck time, then the temperature uncertainty $\delta T$ for this measurement equals the Planck temperature.

The nontrivial BH microstate degeneracy starts at 3 vertices \cite{29} of the Voronoi triangulation corresponding to the Delaunay triangulated $\pi$-bit BH with 4 vertices (Fig.~\ref{Fig_03}(a)). This is interesting in the context of the equipartition theorem for an atom in a monatomic ideal gas, that --- assuming that the atom exhibits $\pi$ degrees (3 bits+) of freedom, instead of three --- can be defined as
\begin{equation}\label{68}
E=\frac{\pi}{2} k_{B} T .
\end{equation}
This form enables the recovery of the exact (not approximate) Unruh \eqref{38} and Hawking \eqref{39} temperature equations using derivations presented in \cite{31} if energy uncertainty $\delta E$ is interpreted in terms of temperature uncertainty $\delta T$. Notably, the equipartition theorem is rigorously proven only for a single degree of freedom (one bit), and the energy \eqref{68} corresponds to the $\pi$-bit BH, which can be seen by substituting BH mass-energy equivalence \eqref{64} along with BH temperature \eqref{11} into the $\pi$-bit equipartition theorem \eqref{68}.

Based on the Planck-Einstein (or Compton wavelength) relation
\begin{equation}\label{69}
\delta E=h \delta v=\frac{2 \pi \hbar c}{\delta \lambda},
\end{equation}
where the uncertainty of energy $\delta E$ is expressed as the uncertainty of wavelength $\delta \lambda$ the HUP \eqref{66} can be expressed as
\begin{equation}\label{70}
\delta \lambda \leq 4 \pi c \delta t \quad(l \leq 4 \pi t) .
\end{equation}
Alternatively, the mass-energy equivalence $\delta E=\delta M c^2$ can be used to replace energy uncertainty with mass uncertainty $\delta M$, and set $c=\delta R / \delta t$ \cite{31} to express HUP \eqref{66} as
\begin{equation}\label{71}
\delta M \delta R \geq \hbar / 2 c .
\end{equation}
On substituting $\delta M=2 \pi \hbar /(c \delta \lambda)$ (Compton wavelength of $\delta M$) into HUP \eqref{71}, we get
\begin{equation}\label{72}
\delta \lambda \leq 4 \pi \delta R \quad(l \leq 2 \pi d),
\end{equation}
which also supports the finding that for $\delta \lambda \geq \ell_{P}$ ($l \geq 1$) \eqref{4}, $d_{BH} \geq 1 /(2 \pi)$ sets a maximum BH Planck temperature \eqref{11} equal to the Planck temperature. Moreover, by substituting $\delta M=\delta R c^{2} / (2 G)$ (Schwarzschild mass corresponding to $\delta R$) into HUP \eqref{71}, we can recover the Planck length \cite{31} as the minimum BH radius, expressed as
\begin{equation}\label{73}
\delta R^{2} \geq \ell_{P}^{2}, \quad c^{2} \delta t^{2} \geq \ell_{P}^{2},
\end{equation}
and hence, the Planck time $t_{P}$ can be considered as the minimum time period $\delta t$ on a BH horizon for one bit.

In this section, we introduced the concept of the disturbing time period $\delta t$ and equated it with time uncertainty. Although we implicitly equated the position uncertainty with the disturbing radius $\delta R$ or BH radius $R_{BH}$ \eqref{26} (as shall be discussed later, they are additive inverses of each other), it is irrelevant in the current context because the uncertainties in HUP are always non-negative. These aspects are further discussed in Section \ref{sec:9}.

Micro-BHs are inherently unstable and prone to collapses \cite{20}. The HUP, BH geometry, and information theory further deliver arguments in support of this claim. In particular, the BH temperature limit \eqref{11} is $d_{BH}=1 /(2 \pi)$, whereas $d_{BH}=1 / \sqrt{\pi}$ provides one Planck area (one bit). At least four vertices, the fundamental requirement for defining the event horizon \eqref{13}, are provided by $d_{BH}=1$. Moreover, the imaginary time period is exhibited by BH with $d_{BH}=\sqrt{2}$ \eqref{103}. Twelve Planck areas (with only one precise diameter) correspond to the BH with $d_{BH}=2$, which is also recovered from the HUP \eqref{73}. Compared to \num{7.24E90} Planck areas of Sagittarius A$^*$, the maximization of Kolmogorov complexity of a BH binary message \eqref{59} comprising 20 Planck areas is much more challenging.

Generally, BH radius $R_{BH}$ is derived as a singularity of the Schwarzschild metric, which yields an exact solution to the Einstein field equations without the cosmological constant $\Lambda$, thereby describing the gravitational field outside a spherical, uncharged, and nonrotating mass. Note that the cosmological constant introduces concepts like dark matter, dark energy, or dark fluid which are redundant in the framework of Verlinde's entropic gravity \cite{23}. More specifically, the Schwarzschild metric has two singularities: at $R=0$ ($R=0$ is always singular in polyspherical coordinates) and at $R=R_{BH}$.
However, $R_{BH}$ is also a solution to the escape velocity threshold, as follows
\begin{equation}\label{74}
v_{E}^{2}=\frac{2 G M}{R} \leq c^{2} \Leftrightarrow \frac{G M}{R_{BH}}=\frac{1}{2} c^{2} .
\end{equation}

By equating variational potential \eqref{50} with binary potential \eqref{54}, we can determine the relation between the mass $M$ and radius $\delta R$. This can be conducted for $\delta \varphi_{k}=-c^{2}$ (it is undefined for $\delta \varphi_{k}=0$ ) resulting in
\begin{equation}\label{75}
\delta R_1=\frac{G M}{c^{2}},
\end{equation}
which represents half the Schwarzschild radius $\delta R_{BH}$ of mass $M$ and equals the Planck length if $M$ is equal to the Planck mass. Therefore, we recovered the orbital velocity threshold with respect to $\delta R_{1}$, as follows
\begin{equation}\label{76}
v_{O}^{2}=\frac{G M}{\delta R_{1}} \leq c^{2},
\end{equation}
which expresses the second (or rather, the first) significant velocity related to mass $M$. This threshold is meaningless in the context of the Schwarzschild metric, as it simply does not encompass orbital velocity (although it encompasses the photon-sphere radius $R_{P}=3 G M / c^{2}$). Orbital velocity is included in the Kerr metric, which is another exact solution to the Einstein field equations without the cosmological constant $\Lambda$ and acts as a generalization of the Schwarzschild metric to the gravitational field prevailing outside a spherical and uncharged mass, rotating with an angular momentum $J$. In contrast, Birkhoff's theorem asserts that a spherically symmetric solution to Einstein's equations in the vacuum must necessarily be static, and the exterior solution must be described by Schwarzschild metric in the absence of the cosmological constant \cite{32}.

Interestingly, for constant $M$, a hypothetical 2 -sphere of radius $\delta R_{1}$ \eqref{75} comprises the same number of bits as given by BH entropy \eqref{44} (using mass discretization \eqref{2})
\begin{equation}\label{77}
\begin{aligned}
&N_{BH}=4 \pi \frac{4 G^{2} M^{2}}{c^{4}} \frac{c^{3}}{\hbar G}=16 \pi m^{2} \\
&N_{\delta R_{1}}=4 \pi \frac{G^{2} M^{2}}{c^{4}} \frac{c^{3}}{\hbar G}=4 \pi m^{2}=\frac{1}{4} N_{BH}.
\end{aligned}
\end{equation}

Furthermore, by equating Verlinde's entropy variation \eqref{46} ($2^{\text{nd}}$ term) with entropy variation \eqref{53} ($2^{\text {nd }}$ term), we can express variational potential \eqref{50} on the entropy variation sphere as a function of the information capacity $N_{A}$ of the sphere and the reduced Compton wavelength of mass $M$, given as
\begin{equation}\label{78}
\delta \varphi=-\frac{4 \pi c^{2}}{N_{A}} \frac{\delta R}{\lambdabar_{M}} .
\end{equation}

Combining variational potential \eqref{78} with variational potential \eqref{50} ($2^{\text{nd}}$ term), we get
\begin{equation}\label{79}
N_{A}=\frac{4 \pi \delta R^{2}}{\ell_{P}^{2}}, \quad \text { and since, } \quad N_{A}=\frac{4 \pi R^{2}}{\ell_{P}^{2}},
\end{equation}
we conclude that both $\delta R$ and $R$ provide the same information capacity $N_{A}$ on the spherical screen $A$. However, $\delta R$ is related to the notion of the time period $\delta t$ \eqref{31}, whereas $R$ is not. Thus, we conclude that $R^{2}=\delta R^{2}$. Additionally, $\delta R$ is an additive inverse of $R$, as discussed in the subsequent section.

\section{Inertial Potential}\label{sec:8}
Here, we introduce the concept of the inertial potential of the active mass $M$
\begin{equation}\label{80}
\phi_{a}=\frac{G M}{R},
\end{equation}
as an additive inverse of this mass gravitational potential \cite{25}, and consider it to be equivalent to the latter. Indeed, the gravitational field and inertia are physically equivalent (equivalence principle), wherein the inertia originates through interactions between all masses in the universe (Mach's principle) on which principles the theory of relativity is based \cite{33}. Potential $\phi$ bounds the notions of mass and space, similar to mass density $\rho$, Schwarzschild radius \eqref{26}, or the uncertainty principle. Normal modulation of gravitational potential \cite{25}, caused by rotating bodies is wrongly interpreted as a gravitational wave understood as a carrier of gravity \cite{szostek}.

Based on the definition of binary potential \eqref{54}, we observed that radius $R$ emerges as irrelevant in inertial potential \eqref{80}. By introducing the gradient of inertial potential \eqref{83} into Unruh temperature \eqref{38} and substituting it along with binary entropy variation \eqref{55} and the radius of the holographic sphere $R^{2}=N_{A} \ell_{P}^{2} / 4 \pi$ into entropic work \eqref{43} we obtain entropic work (using Equation \eqref{62} in the last term), given as \begin{equation}\label{81}
\begin{aligned}
&F \circ \delta R=T_{a} \cdot \delta S=\frac{\hbar}{2 \pi c k_{B}}\left(-\frac{G M}{R^{2}}\right) \cdot\left(\frac{1}{2} k_{B} N_{1}\right)=\\
&=-\frac{\hbar G}{4 \pi c} M \frac{4 \pi}{N_{A} \ell_{P}^{2}} N_{1}=-\frac{N_{1}}{N_{A}} M c^{2}=M \delta \varphi.
\end{aligned}
\end{equation}
Therefore, mass $M$ represents the proportionality constant, similar to that in Isaac Newton's second law of motion \eqref{47}, and the actual radius $R$ of our toy universe, portrayed in Fig.~\ref{Fig_06}, is irrelevant to our considerations.

The supermassive BH Sagittarius A$^*$ has an estimated mass $M_{BH} \approx \num{8.26E36}$ kg, yielding diameter $D_{BH} \approx \num{2.45E10}$ m. The BH surface gravity can be expressed as\footnote{$g = 1/d$, where $g$ is the Planck acceleration multiplier.}
\begin{equation}\label{82}
g_{BH}=\frac{G M_{BH}}{R_{BH}^{2}}=\frac{c^{2}}{D_{BH}},
\end{equation}
which is independent of the gravitational constant $G$.
This is an extremely large value, even for supermassive BHs. For BHs listed in Table~\ref{table1}, $g_{BH}$ displays an order of magnitude of the Planck acceleration $a_{P}=\sqrt{ }\left(c^{7} / h G\right)=\num{5.56E51}$ m/s$^2$, attained by the $\pi$-bit BH. For Sagittarius A$^*$, this is still $g_{BH} \approx \num{3.66E6}$ m/s$^2$. A BH with $g_{BH}$ equal to the Earth's surface gravity (9.81 m/s$^2$) would require a diameter of $D_{BH} \approx \num{9.16E15}$ m, which is slightly less than one light year.

Such large black holes are not observed in nature.

Thus, any inertial acceleration of an object heavier than the threshold of distinguishability \eqref{6} can be considered to be lower than a BH surface gravity. The inertial acceleration $a$ denotes the gradient of inertial potential \eqref{80}, such that
\begin{equation}\label{83}
a=\nabla_{R} \phi_{a}=-\frac{G M}{R^{2}}<\frac{c^{2}}{2 R_{BH}},
\end{equation}
thereby indicating that inertial potential \eqref{80} of such a mass at $R_{BH}=R$ must be greater than BH potential \eqref{27} of $-c^2/2$
\begin{equation}\label{84}
\phi_{a}=\frac{G M}{R}>-\frac{1}{2} c^{2} .
\end{equation}

This is an important finding since it signifies that inertial potential \eqref{80} of such a mass always describes a non-equilibrium situation, where the equilibrium is defined by the BH thermodynamic equilibrium \eqref{56} and Laplace's \eqref{28} or Poisson's \eqref{29} equation.

Similarly, if we presume that variational potential \eqref{50} should be higher than BH potential \eqref{27}
\begin{equation}\label{85}
\delta \varphi \doteq-\frac{G M}{\delta R}>-\frac{1}{2} c^{2},
\end{equation}
since in this case mass $M$ is away from the equilibrium \eqref{56}, we obtain
\begin{equation}\label{86}
\delta R>\frac{2 G M}{c^{2}},
\end{equation}
which indicates that the disturbing radius $\delta R$, wiggling mass $M$ against Poisson's equation \eqref{29} must be larger than the Schwarzschild radius \eqref{26} of mass $M$. If $M$ forms a BH, it cannot be wiggled, as this would invalidate the BH equilibrium \eqref{56} condition and Bekenstein's threshold \eqref{44} on a BH entropy. The disturbing radius $\delta R$ should also be larger \eqref{73} than the Planck length.

\begin{figure}[ht]
\includegraphics[width=0.5\textwidth]{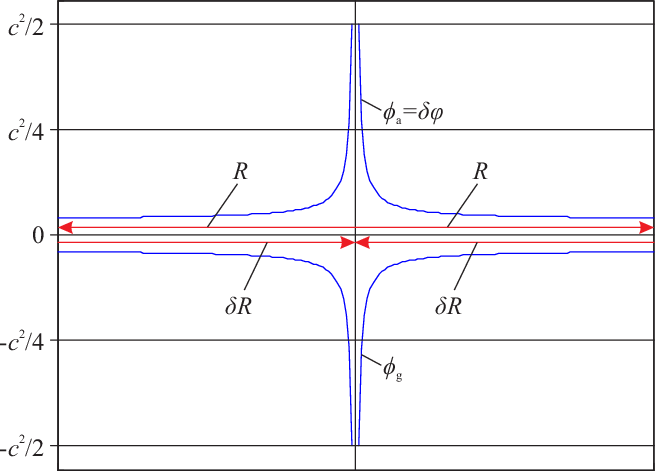}
\caption{\label{Fig_07}  Gravitational, inertial, and variational potentials (not to scale).}
\end{figure}

Based on Equations \eqref{79}, \eqref{81}, \eqref{84}, and \eqref{85}, we can conclude that variational potential \eqref{50} represents gravitational potential corresponding to inertial potential \eqref{80}, i.e.
\begin{equation}\label{87}
\delta R \geq \ell_{P}, \quad \delta R^{2}=R^{2}, \quad \delta R=-R, \quad \delta \varphi=\phi_{a},
\end{equation}
as shown in Fig.~\ref{Fig_07}.
Variational potential \eqref{78} induced by wiggling the location of the test mass $M$ must be equal to the BH potential of $-c^{2} / 2$, assuming that the test mass $M$ could be wiggled on a BH horizon. Thus, using $\delta R=-R_{BH}$ \eqref{87},
\begin{equation}\label{88}
\delta \varphi=-\frac{4 \pi c^{2}}{N_{A}} \frac{\delta R}{\lambdabar_{M}}\myeq{}\frac{4 \pi c^{2}}{N_{BH}} \frac{R_{BH}}{\lambdabar_{M}} \doteq-\frac{c^{2}}{2},
\end{equation}
and a simple calculation (cf. Appendix \ref{app:8}) reveals that the reduced Compton wavelength of the active mass $M$ in Equation \eqref{88} is the additive inverse of the reduced BH Compton wavelength \eqref{9}, corresponding to BH mass required to be emitted to ensure BH collapse \eqref{20}, given as
\begin{equation}\label{89}
\lambdabar_{M}\myeq{}-\frac{2 \ell_{P}^{2}}{R_{BH}} = -\lambdabar_{BH} .
\end{equation}
This is not surprising, considering $M$ exhibits an ultimately attained equilibrium \eqref{56}, and hence, becomes a portion of BH. However, this is negative because $M$ was introduced to $M_{BH}$ in a direction opposite to its radius.

In a subsequent section, we shall investigate the meaning of time period $\delta t$ and the distance $\delta L$ tangential to the holographic sphere $A$.

\section{Imaginary Time and Kinematics on Holographic Spheres}\label{sec:9}
Let us postulate the velocity $v_{R}$ as normal to a holographic sphere, and thus, unobservable. Evidently, only one axis is normal to this sphere at each Planck area, whereas the number of tangential axes is infinite. Let us further postulate that this unobservable velocity $v_{R}$ is bound with the classical observable velocity
\begin{equation}\label{90}
v=\frac{\delta L}{\delta t},
\end{equation}
where $\delta L$ denotes a certain length, orthogonal to the disturbing radius $\delta R$ (and by Equation \eqref{87} also orthogonal to this sphere radius $R$) based on the Pythagorean relation
\begin{equation}\label{91}
v^{2}+v_{R}^{2}=c^{2}.
\end{equation}

\begin{figure}[ht]
\includegraphics[width=0.5\textwidth]{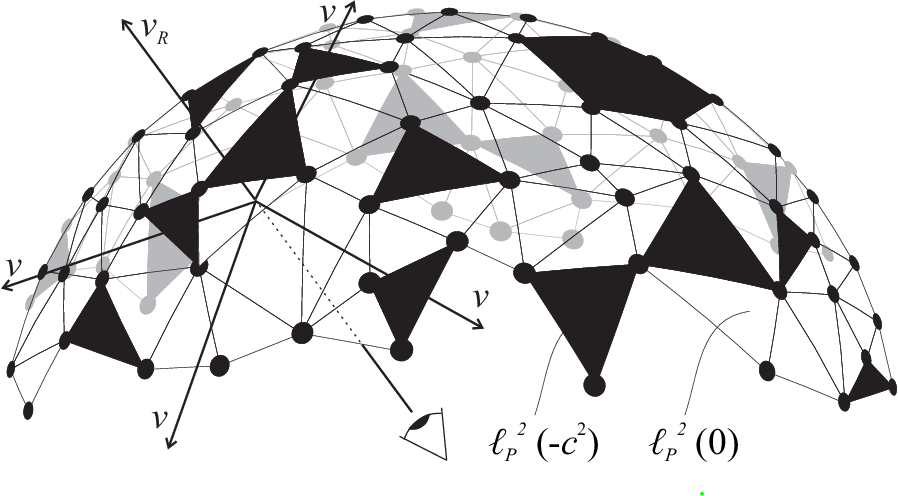}
\caption{\label{Fig_08} Simplified illustration of visual perception of movement on a holographic patterned sphere (not to scale).}
\end{figure}

We assume that $\delta L$ is observable as moving on the screen owing to the varying binary potentials \eqref{54}, as schematically shown in Fig.~\ref{Fig_08}. Contrary to BHs not having interiors, each observer is bounded within a 2-dimensional sphere $A$, with two spatial dimensions encoding its information capacity $N_{A}$ with $\left\lfloor N_{A}\right\rfloor$ Planck area spherical triangles swapping in imaginary time, whereas $0 i=0$ denotes the present moment of perception (observation) when an active Planck area can become inactive, and vice versa. However, we do not investigate the precise mechanism of this process. It is known, for example, that the photoisomerization of a photon in a cone cell of the eye cascades signal transduction.

Based on the postulated velocity relation \eqref{91}, the Lorentz factor can be derived as
\begin{equation}\label{92}
\gamma=\left(1-\frac{v^{2}}{c^{2}}\right)^{-\frac{1}{2}}=\frac{c}{v_{R}},
\end{equation}
the Lorentz contraction (squared) becomes $\left(0 \leq \delta L^{2} \leq \delta L_{0}^{2}\right)$
\begin{equation}\label{93}
\delta L^{2}=\delta L_{0}^{2}\left(1-\frac{v^{2}}{c^{2}}\right)=\delta L_{0}^{2} \frac{v_{R}^{2}}{c^{2}},
\end{equation}
and time dilation (squared) becomes $\left(0 \leq \delta t_{0}{ }^{2} \leq \delta t^{2}\right)$
\begin{equation}\label{94}
\delta t^{2}=\delta t_{0}^{2} /\left(1-\frac{v^{2}}{c^{2}}\right)=\delta t_{0}^{2} \frac{c^{2}}{v_{R}^{2}} .
\end{equation}
Overall, the simple forms of these equations endorse the validity of the postulated velocity relation \eqref{91}.

According to Equations \eqref{38}, \eqref{39}, and \eqref{40}, the observable acceleration
\begin{equation}\label{95}
a=\frac{\delta R}{\delta t^{2}},
\end{equation}
is perpendicular to the holographic sphere. Additionally, the application of the disturbing radius $\delta R$ and disturbing time period $\delta t$ in observable acceleration $a$ is justified by our previous conclusion \eqref{84} that any inertial acceleration introduces a nonequilibrium condition.

Let us then postulate that this observable acceleration $a$ is, similarly to velocity \eqref{91}, bound with a certain unobservable orthogonal acceleration $a_{T}$, which is tangential to the holographic sphere at a given Planck area, based on a similar Pythagorean relation
\begin{equation}\label{96}
a_{T}^{2}+a^{2}=a_{P}^{2}=\frac{c^{2}}{t_{P}^{2}}=\frac{\ell_{P}^{2}}{t_{P}^{4}} .
\end{equation}

The observable acceleration \eqref{95} can be expressed using Unruh temperature \eqref{38} as
\begin{equation}\label{97}
a=\frac{2 \pi c k_{B}}{\hbar} T_{a}=2 \pi a_{P} \frac{T_{a}}{T_{P}} .
\end{equation}
Substituting Equation \eqref{97} into acceleration relation \eqref{96} we get
\begin{equation}\label{98}
a_{T}=a_{P} \sqrt{1-4 \pi^{2} \tau^{2}},
\end{equation}
where
\begin{equation}\label{99}
\tau \doteq \frac{T_{a}}{T_{P}}=\frac{1}{2 \pi} \frac{a}{a_{P}} \leq 1 \quad \tau \in \mathbb{R},
\end{equation}
denotes the ratio of the Unruh temperature or acceleration to the Planck temperature or acceleration. If $\tau=1 /(2 \pi)$, $a_{T}$ vanishes in acceleration relation \eqref{98} and only radial acceleration $a$, equal to the Planck acceleration, $a_{P}$ exists on the holographic sphere having diameter of a $\pi$-bit BH shown in Fig.~\ref{Fig_03}(a). If $\tau=0$ (absolute zero), then $a_{T}=a_{P}$ and the radial acceleration $a$ vanishes in \eqref{96}.

At BH horizon $T_a=T_{BH}$ \eqref{11}, thus $\tau_{BH}=1 /(2 \pi d)$ and acceleration relation \eqref{98} becomes
\begin{equation}\label{100}
a_{T}\myeq a_{P} \sqrt{1-\frac{1}{d_{BH}^{2}}},
\end{equation}
which is imaginary if $d_{BH}^{2}<1$. The real values of $a_{T}$ are first produced by  the $\pi$-bit BH ($d_{BH}=1$) with surface gravity \eqref{82} equaling to the Planck acceleration, which also validates postulated acceleration relation \eqref{96}.

Moreover, Unruh temperature \eqref{38} can be expressed as
\begin{equation}\label{101}
T_{a}=\frac{\hbar}{2 \pi c k_{B}} \frac{\delta R}{\delta t^{2}}=\frac{r_{\delta}}{2 \pi t^{2}} T_{P},
\end{equation}
where $\delta R=r_{\delta} \ell_{P}$, $\delta t=t\cdot t_{P}$, and $r_{\delta}, t \in \mathbb{R}$ denote the Planck length and time multipliers in $\delta R$ and $\delta t$. At a BH horizon, this temperature must be equal to BH temperature \eqref{11}, which results in
\begin{equation}\label{102}
\frac{r_{\delta}}{2 \pi t^{2}}\myeq\frac{1}{2 \pi d_{BH}},\;\;\; r_{\delta}\left(-2 r_{\delta}\right)\myeq t^{2},\;\;\; \frac{\delta R^{2}}{\delta t^{2}}\myeq-\frac{1}{2} c^{2},
\end{equation}
where we used $r_{\delta}=-r_{BH}$ \eqref{87}, so the sign is reversed. This is equal to the BH potential \eqref{27}. 

Since $\delta R$ is real, we conclude that at BH horizon time period $\delta t$ is imaginary
\begin{equation}\label{103}
\frac{\delta R}{\delta t}\myeq\frac{i}{\sqrt{2}} c,\;\;\; \delta t\myeq\frac{i D_{BH}}{c \sqrt{2}},
\end{equation}
and thus, that the unobservable velocity $v_{R}$ is also imaginary and can be expressed as real
\begin{equation}\label{104}
v_{R}=\frac{\delta R}{\delta t_{R}}=\frac{\delta R}{i \delta t},
\end{equation}
upon the introduction of the complementary time period $\delta t_{R}$ related to the \emph{classical} time period $\delta t$ based on integral powers of the imaginary unit $i$
\begin{equation}\label{105}
\delta t_{R}=i \delta t,\;\;\; \delta t_{R}^{2}=-\delta t^{2},\;\;\; \delta t_{R}^{3}=-i \delta t^{3},\;\;\; \delta t_{R}^{4}=\delta t^{4}.
\end{equation}

Thus, at the BH horizon, Equations \eqref{103} and \eqref{104} can be combined to derive
\begin{equation}\label{106}
v_{R}^{2}=\frac{\delta R^{2}}{\delta t_{R}^{2}}\myeq\frac{1}{2} c^{2},
\end{equation}
which represents the inverse of a BH potential \eqref{27}, and using velocity relation \eqref{91} provides the observable velocity of the same magnitude
\begin{equation}\label{107}
v^{2}=\frac{\delta L^{2}}{\delta t^{2}}\myeq\frac{1}{2} c^{2}.
\end{equation}

Substituting $\delta t=i t_{P}$, the shortest theoretically measurable time period, into Equation \eqref{103} yields $d_{BH}=\sqrt{2}$, which is the minimum BH diameter multiplier allowing for the notion of (imaginary) time.

Furthermore, Equations \eqref{102} and \eqref{103} yield the acceleration as an additive inverse of the BH surface gravity \eqref{82}
\begin{equation}\label{108}
a=\frac{\delta R}{\delta t^{2}} \myeq-\frac{c^{2}}{2 \delta R}=\frac{i c}{\sqrt{2} \delta t},
\end{equation}
which is congruent with $R$ to $\delta R$ relation \eqref{87}.

Using unobservable velocity \eqref{104}, time dilation \eqref{94} can be expressed as
\begin{equation}\label{109}
\frac{\delta R^{2}}{\delta t_{0}^{2}}=-c^{2},
\end{equation}
which is independent both on the time period $\delta t$ measured in a moving inertial frame of reference and the velocity $v$ of this frame. The velocity relation \eqref{91} and Lorentz contraction \eqref{93} results in (cf. Appendix \ref{app:12})
\begin{equation}\label{110}
\frac{\delta L_{0}^{2}}{\delta L^{2}}+\frac{\delta L^{2}}{\delta R^{2}}=1,
\end{equation}
which is neither time period nor velocity dependent.

Let us now postulate that the unobservable acceleration has the form of
\begin{equation}\label{111}
a_{T}=\frac{\delta L}{\delta t_{R}^{2}} .
\end{equation}

The current conclusions concerning the velocities \eqref{90} and \eqref{104} and accelerations \eqref{95} and \eqref{111} are implied by Mach's principle: observable velocity provides information on a holographic sphere, whereas observable acceleration acts on the holographic sphere and is originated in an interaction between all masses in the universe. To illustrate this principle, we quote Steven Weinberg: 

\begin{center}
''There is a simple experiment that anyone can perform on a starry night, to clarify the issues raised by Mach's principle. First, stand still, and let your arms hang loose at your sides. Observe the stars are more or less unmoving, and that your arms hang more or less straight down. Then pirouette. The stars will seem to rotate around the zenith, and at the same time your arms will be drawn upward'' \cite{34}.
\end{center}

The questions that arise are, why do one's arms hang loose when the stars are still, and why they would be drawn upward when the stars rotate? The most important factor for our definitions of the observable velocity \eqref{90}, acting tangential to the holographic sphere, and the observable acceleration \eqref{95}, acting perpendicularly to it, is that the \emph{stars} simulated to whirl around a person standing in the center of a planetarium would not draw this person's arms upward.

The velocity \eqref{91} and acceleration \eqref{96} relations can now be expressed in terms of $\delta t$, using time periods relation \eqref{105} as
\begin{equation}\label{112}
\frac{\delta L^{2}}{\delta t^{2}}+\frac{\delta R^{2}}{\delta t_{R}^{2}}=\frac{\delta L^{2}}{\delta t^{2}}-\frac{\delta R^{2}}{\delta t^{2}}=c^{2},
\end{equation}
\begin{equation}\label{113}
\frac{\delta L^{2}}{\delta t_{R}^{4}}+\frac{\delta R^{2}}{\delta t^{4}}=\frac{\delta L^{2}}{\delta t^{4}}+\frac{\delta R^{2}}{\delta t^{4}}=a_{P}^{2}=\frac{c^{2}}{t_{P}^{2}}.
\end{equation}
Dividing Equation \eqref{112} by $\delta t^{2}$ and adding it to Equation \eqref{113} we arrive at the relation
\begin{equation}\label{114}
\delta L^{2}=\frac{1}{2} \delta t^{2}\left(a_{P}^{2} \delta t^{2}+c^{2}\right),
\end{equation}
bounding $\delta L$ with $\delta t$. On the other hand, velocity \eqref{91} and acceleration \eqref{96} relations can be expressed in terms of $\delta t_{R}$, yielding a similar bound between $\delta R$ with $\delta t$
\begin{equation}\label{115}
\delta R^{2}=\frac{1}{2} \delta t^{2}\left(a_{P}^{2} \delta t^{2}-c^{2}\right).
\end{equation}
Adding Equation \eqref{114} to Equation \eqref{115} yields
\begin{equation}\label{116}
\delta L^{2}+\delta R^{2}=a_{P}^{2} \delta t^{4}.
\end{equation}
Discretized by the Planck length and time for natural multipliers $r$, $t$, $a$, Equation \eqref{116} represents the OEIS A349078 sequence.

\begin{figure}[ht]
\includegraphics[width=0.5\textwidth]{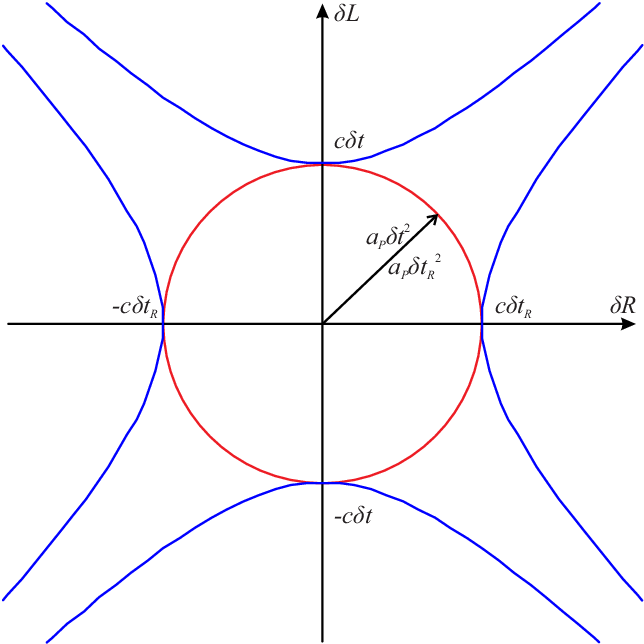}
\caption{\label{Fig_09}  Pythagorean (red) and hyperbolic (blue) relations between $\delta L$ perpendicular to the disturbing radius $\delta R$.}
\end{figure}

We observe that $\delta L$ and $\delta R$ are bound by the Pythagorean (circle) relation \eqref{116} and also by the hyperbolic relation
\begin{equation}\label{117}
\delta L^{2}-\delta R^{2}=c^{2} \delta t^{2},
\end{equation}
as shown in Fig.~\ref{Fig_09}.

Dividing Equations \eqref{114} and \eqref{115} with $\delta t^{2} \neq 0$, we get
\begin{equation}\label{118}
\frac{\delta L^{2}}{\delta t^{2}}=\frac{1}{2} c^2\left(t^{2}+1\right)=v^{2} \leq c^2,
\end{equation}
and
\begin{equation}\label{119}
\frac{\delta R^2}{\delta t^{2}}=\frac{1}{2} c^{2}\left(t^{2}-1\right)=-v_{R}^{2} \leq c^{2} .
\end{equation}
Considering, no velocity exceeds the speed of light $c$, the bounds are set on inequalities \eqref{118} and \eqref{119}, as listed in Table~\ref{table4}. Based on BH bounds \eqref{106} and \eqref{107}, we can conclude that
\begin{equation}\label{120}
0 \leq \frac{\delta L^{2}}{\delta t^{2}} \leq \frac{1}{2} c^{2},\;\;\; 0 \leq \frac{\delta R^{2}}{\delta t^{2}} \leq \frac{1}{2} c^{2}
\end{equation}

\begin{table}
\caption{\label{table4} Characteristic Planck time multipliers.}
\begin{ruledtabular}
\begin{tabular}{|c|ccc|}
 & 0 & $c^2 / 2$ & $c^2$ \\
 \hline
$\delta L^2 / \delta t^2$ & $t^2=-1$ & $t^2=0$ & $t^2=1$ \\
$\delta R^2 / \delta t^2$ & $t^2=1$ & $t^2=2$ & $t^2=3$
\end{tabular}
\end{ruledtabular}
\end{table}
On substituting Equations \eqref{118} and \eqref{119} into the orbiting condition $v_{O}^{2} \leq v^{2} \leq v_{E}^{2}$, with thresholds \eqref{76} and \eqref{74} we get
\begin{equation}\label{121}
\frac{R_{BH}}{R} \leq t^{2}+1 \leq \frac{2 R_{BH}}{R},
\end{equation}
and
\begin{equation}\label{122}
\frac{R_{BH}}{R} \leq t^{2}-1 \leq \frac{2 R_{BH}}{R} .
\end{equation}
These are interesting results. Our toy universe, shown in Fig.~\ref{Fig_06}, contains a black hole in its center. But it is not necessary for the above relations to be valid. The Schwarzschild radius of the Sun (\num{1.99E30} kg), for example, amounts to $R_{BH}=\num{2.95E3}$ m.

At BH horizon $R=R_{BH}$, and hence, $0 \leq t^{2} \leq 1$ and $2 \leq t^{2} \leq 3$ in Equations \eqref{121} and \eqref{122}, which corresponds to the $c^{2} / 2$ to $c^{2}$ range in Table~\ref{table4}. 

Notably, de Broglie wavelength (of a matter wave) corresponds to the Compton wavelength of that wave for $v^2 = c^2/2$.

Furthermore, we observe that the range 0 to $c^2/2$ in Table~\ref{table4} is achieved by $0 \leq t^{2}+1 \leq 1\left(-1 \leq t^{2} \leq 0\right)$ and $0 \leq t^{2}-1 \leq 1\left(1 \leq t^{2} \leq 2\right)$ with a gap $0 \leq t^{2} \leq 1$. Comparing this with Equations \eqref{121} and \eqref{122}, we can conclude that the maximum significant BH radius equals $2 R_{BH}$. Notably, there are no celestial objects other than BHs\footnote{Even white dwarfs (such as Sirius B; $M=\num{2.02E30}$ kg, $R=\num{5.84E6}$ m $\rightarrow \phi_g=\num{-2.31E13}$ m$^2$/s$^2$) and neutron stars (such as PSR J 0740+6620; $M=\num{4.14E30}$ kg, $R=\num{1.24E4}$ m $\rightarrow \phi_{g}=\num{-2.24E16}$ m$^2$/s$^2$) provide potential lower (up to modulus) than BHs, with the latter being close to $-c^2/4=\num{-2.25E16}$ m$^2$/s$^2$.} having modulus of potential larger than $\left|-c^2/4\right|$. For a constant $M_{BH}=R_{BH} c^2/(2G)$ and $R_{MAX}=2R_{BH}$ the Bekenstein bound \cite{18} yields twice BH entropy \eqref{44} ($M_{BH}$ is bounded by $R_{BH}$ but $R$ varies)
\begin{equation}\label{123}
\begin{aligned}
S & \leq \frac{2 \pi k_{B} c R_{M A X} M_{BH}}{\hbar}=2 \pi k_{B} r_{M A X} m_{BH}=\\
&=\frac{2 \pi k_{B} c 2 R_{BH}}{\hbar} \frac{R_{BH} c^{2}}{2 G}=\frac{1}{2} k_{B} N_{BH}=k_{B} N_{1}
\end{aligned}
\end{equation}
which also according to Equation \eqref{56} equals to the number of active Planck areas on a BH horizon. 

We can express the radius $R_{k}$ of a holographic sphere of a constant mass $M=m m_{P}$, $m \in \mathbb{R}$ as a function of $GM/c^2$ multiplier of $k \in \mathbb{R}$, as
\begin{equation}\label{124}
R_{k}=k \frac{G M}{c^{2}}=k m \ell_{P},
\end{equation}
which corresponds to the Schwarzschild radius \eqref{26} of mass $M$ for $k=2$. This yields gravitational potential \eqref{25} $\phi_g\left(R_{k}\right) = -c^2/k$ and the number of the active Planck areas \eqref{62} at the holographic sphere of radius $R_k$ equal to $N_1=N_A/k$. Therefore, there are fewer active Planck areas \emph{outside} the BH horizon of mass $M$ than $N_1=N_A/2$.

Substituting $G$ from $\ell_{P}^{2}=\hbar G/c^3$ into Equation \eqref{124} we get
\begin{equation}\label{125}
Mc \frac{\ell_{P}^{2}}{R_{k}}=\frac{\hbar}{k},
\end{equation}
which for $k=2$ corresponds to HUP bound \eqref{65}. Parameters of the holographic spheres of a constant mass $M$ are listed in Table~\ref{table5}, while Fig.~\ref{Fig_10} shows their binary Shannon entropy \eqref{60} (its properties are discussed in Appendix \ref{app:14})
\begin{equation}\label{126}
H(k)=\ln (k)-\frac{k-1}{k} \ln (k-1) .
\end{equation}

\begin{figure}[ht]
\includegraphics[width=0.5\textwidth]{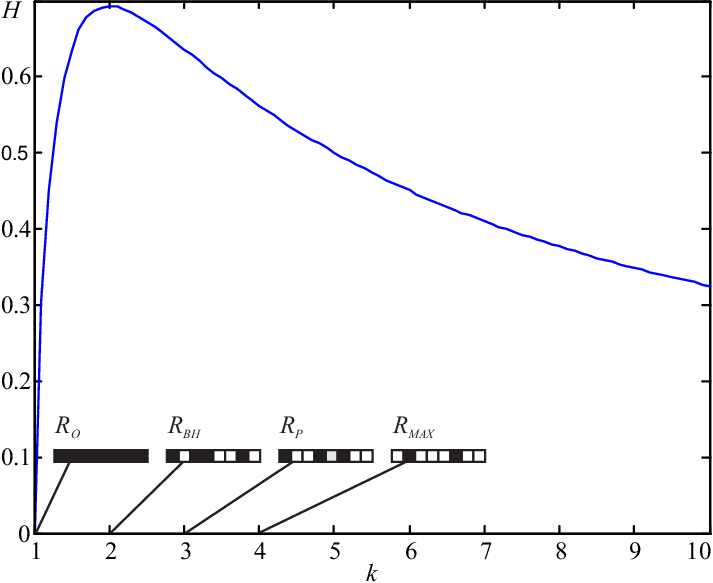}
\caption{\label{Fig_10} Shannon entropy (in nats) of holographic spheres, as a function of $GM/c^2$ multiplier $k$; $M = \text{const}$.}
\end{figure}

Substituting $\delta R^{2}$ from time dilation \eqref{109} into Equation \eqref{119}, we get
\begin{equation}\label{127}
t^{4}-t^{2}-2 t_{0}^{2}=0,
\end{equation}
where the proper time is $\delta t_{0}=i t_{0} t_{P}$ and $t_{0} \in \mathbb{R}$ represents the Planck time multiplier. Subsequent $t_0^2$ from a set of triangular numbers $(0,1,3,6, \ldots$; OEIS A000217) substituted into $t^2=(1 \pm \sqrt{1+8 t_0^2})/2$ from Equation \eqref{127} yield subsequent nonpositive $t^{2}=0,-1,-2,-3, \ldots$ and positive $t^{2}=1,2,3, \ldots$ squares of the Planck time multipliers.

An outstanding property of Boolean $\{0,1\}^{N}$ address space is that the mean Hamming distance between any address $a_{m}$ and all the other addresses (including $a_{m}$) is $N / 2$, whereas the variation is $N / 4$ \cite{35}. If an $\{N\}$-cube \cite{3} was inscribed in the closed $N$-ball most of its vertices would lay at or near the equator, whereas the unit of the active Planck area $\left(-c^{2}\right)$ suggests a particular orientation of BH address space, with vertex $000 \ldots$ representing one pole and vertex $111 \ldots\left[-c^{2}\right]$ representing the other pole. This indicates a link between diameter-only dependent Planck BH blackbody spectral radiance \eqref{130}, \eqref{131} and the corresponding patternless BH binary message \eqref{56}. Interestingly, the fraction of available BH horizon arrangements comprising $N_1=N_{BH}/2$ active Planck areas with binary potential \eqref{54} equal to $-c^2$ decreases from 50 $\%$ (2-bit BH) to zero with $N_{BH}$ approaching infinity. Using Stirling's approximation for large $N_{BH}$ it amounts
\begin{equation}\label{128}
\left(\begin{array}{c}
N_{BH} \\
N_{1}
\end{array}\right) / 2^{N_{BH}} \approx \sqrt{\frac{2}{\pi N_{BH}}}=\frac{\sqrt{2}}{\pi d_{BH}} .
\end{equation}
A relation between a bit of a binary message \eqref{59} and a wavelength multiplier $l$ requires further research. It is certainly easier to maintain \emph{patternlessness} of a longer binary message, which is likely to be responsible for a flatter spectral radiance of larger BHs, such as Sagittarius A$^*$ (cf. Appendix \ref{app:13}). Neutron stars and white dwarfs supported against collapse owing to Pauli exclusion principle, also emit blackbody radiation. Therefore, they also generate patternless binary messages. Perhaps, just one spherical side of BHs hints at their anyonic nature, whereas neutron stars and white dwarfs exhibit fermionic natures.
\begin{table*}
\caption{\label{table5} Size $(R, D, d)$, potential $\left(\phi_{g}\right)$, information capacity $\left(N_A\right)$, no. and probability of occurrence of active Planck areas $\left(N_1\right)$ within all Planck areas $\left(N_A\right)$, Bekenstein bound $\left(S / k_B\right)$, and Shannon entropy $(H)$ of holographic spheres for a constant mass $M=m\,m_D, m \in \mathbb{R}, k \in \mathbb{R}$.}
\begin{ruledtabular}
\begin{tabular}{|l|l|l|l|r|r|r|}
$R$ & $R_{M I N}=0$ & $R_O=m \ell_P$ & $R_{BH}=2 m \ell_P$ & $R_P=3 m \ell_P$ & $R_{M A X}=4 m \ell_P$ & $R_k=k m \ell_P$ \\
$D$ & $D_{M I N}=0$ & $D_O=2 m \ell_P$ & $D_{BH}=4 m \ell_P$ & $D_P=6 m \ell_P$ & $D_{M A X}=8 m \ell_P$ & $D_k=2 k m \ell_P$ \\
$\phi_g=-G M / R$ & NaN & $-c^2$ & $-c^2 / 2$ & $-c^2 / 3$ & $-c^2 / 4$ & $-c^2 / k$ \\
$d=D / \ell_P$ & 0 & $2 m$ & $4 m$ & $6 m$ & $8 m$ & $2 k m$ \\
$N_A=\pi d^2$ & 0 & $4 \pi m^2$ & $16 \pi m^2=N_{BH}$ & $36 \pi m^2$ & $64 \pi m^2$ & $4 \pi k^2 m^2$ \\
$N_1=-\phi_g N_A / c^2$ & NaN & $4 \pi m^2=N_{BH} / 4$ & $8 \pi m^2=N_{BH} / 2$ & $12 \pi m^2=3 N_{BH} / 4$ & $16 \pi m^2=N_{BH}$ & $4 \pi k m^2$ \\
$p_1=N_1 / N_A$ & NaN & 1 & $1 / 2$ & $1 / 3$ & $1 / 4$ & $1 / k$ \\
$S / k_B$ (rel. to $M$) & 0 & $N_{BH} / 8=N_1 / 4$ & $N_{BH} / 4=N_1 / 2$ & $3 N_{BH} / 8=3 N_1 / 4$ & $N_{BH} / 2=N_1$ & $k N_{BH} / 8=2 \pi k m^2$ \\
$H\left(N_1 / N_A\right)$ & NaN & 0 & $\ln (2)$ & $\ln (3)-2 \ln (2) / 3$ & $\ln (4)-3 \ln (3) / 4$ & $\ln (k)-(k-1) \ln (k-1) / k$ \\
\end{tabular}
\end{ruledtabular}
\end{table*}

These considerations are not surprising. We observe and model a BH as a 2-sphere, from which the electromagnetic radiation cannot escape. However, in the context of the graph of nature, the BH is a patternless binary message. Both $\delta t$ and $\delta t_{R}$ are imaginary and become real $(0 i=0)$ only at the moment of interaction/observation of the Planck area $\ell_{P}^{2}$ at the holographic sphere.

\begin{figure}[ht]
\includegraphics[width=0.5\textwidth]{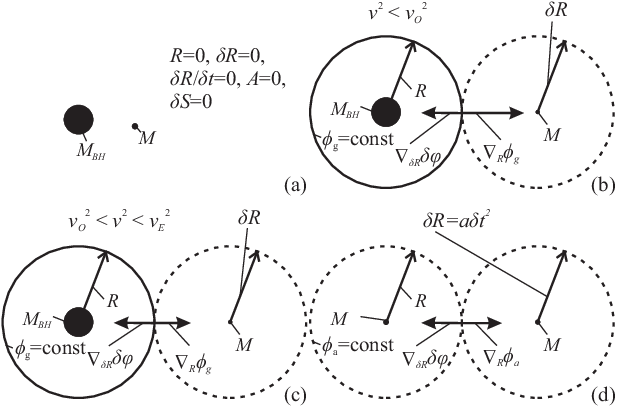}
\caption{\label{Fig_11}   Entropic work of gravity and inertia.
(a) Free fall ($\delta R = 0$, $\delta R/\delta t = 0$, $A = 0$),
(b) spiral trajectory fall $\nu < \nu_O$, (c) stable orbit $\nu_O < \nu < \nu_E$, and
(d) inertial acceleration (not to scale). Gradient sphere (left); entropy variation sphere (right).}
\end{figure}

Certain dynamics scenarios can be contemplated in the simplified model of a spherical universe with two masses (Fig.~\ref{Fig_06}), as shown in Fig.~\ref{Fig_11}. The scenarios involving more than two bodies are not discussed, as they are irrelevant. In any given moment of perception, only two bodies interact with each other through Planck area $\ell_{P}^{2}$ in the Planck time $t_{P}$, and exchange the information in units of 0 or $-c^{2}$ according to the second law of Newton \eqref{47} and Newton's law of gravity \eqref{48}.

The mass $M$ in a state of a timeless free fall on $M_{BH}$ is shown in Fig.~\ref{Fig_11}(a), where $\delta R=0$. Therefore, no entropic work occurs, and the system is entirely described by Poisson's equation \eqref{29} with the inhomogeneity of the mass density included in the $4 \pi G \rho$ term. This equation is independent of time or the variations of other quantities (position, velocity) with time. In particular, variational potential $\delta \varphi$ \eqref{50} is infinite and $A=0$ for vanishing $\delta R$; therefore, the entropy variation   $\delta S$ \eqref{41} is zero, such as that in the limiting case of the binary Shannon entropy (for $p_0=0$ or $p_1=0$ ), where $0 \ln (0)$ is defined as zero.

Entropic work \eqref{43} is shown in Fig.~\ref{Fig_11}(b-d). The notion of time emerges from instantaneous entropy variations \eqref{41} induced by instantaneous variations of $\delta R$ in time $\delta t$.

As shown in Fig.~\ref{Fig_11}(b), the mass $M$ moves along a spiral trajectory towards $M_{BH}$. Obviously, the geometry of this trajectory is not described in Fig.~\ref{Fig_11}(b). Although entropy variation $\delta S$ \eqref{41} is defined and nonvanishing ($\delta R \neq 0$ and $\delta R / \delta t \neq 0$), the velocity of $M$ is below the orbital velocity $v_{O}$ defined by $M_{BH}$ and $R$ ($M \ll M_{BH}$). Consequently, a holographic 2-sphere $A$ emerges both as the sphere of gravitational potential \eqref{25} and as entropy variation sphere \eqref{50}. Subsequently, as time progresses, the system aims toward the equilibrium \eqref{56}, $N_1(t) \to N_{BH}(t)/2$.

As shown in Fig.~\ref{Fig_11}(c), the velocity of $M$ is greater than the orbital velocity $v_{O}$ and lower than the escape velocity $v_{E}$. Therefore, $M$ orbits around $M_{BH}$. If the entropic work is being done, $\delta R \neq 0$ and $\delta R / \delta t \neq 0$, which implies fluctuations in the instantaneous velocity of an orbiting body ranging between the orbital and escape velocities. In this case, $N_{1}(t)<N_{A}(t) / 2$ and the structure of $M$ and $M_{BH}$ is dissipative \cite{12}.

As shown in Fig.~\ref{Fig_11}(d), $M$ accelerates with an acceleration of $a=\delta R / \delta t^{2}$ in the absence of background mass. The gradient \eqref{83} of the inertial potential \eqref{80} provides a conceptual radius $R$, which defines an emergent instantaneous holographic sphere $A$, whereas $\delta R$ provides the entropy variation sphere \eqref{50}. Eventually, the conceptual $R$ is eliminated in the entropic work \eqref{81}, thereby leaving only the conceptual $2^{\text {nd }}$ mass $M$. The gradient \eqref{83} of the inertial potential \eqref{80} on $A$ is normal to both sides of $A$. The area of the holographic sphere $A$ increases as the inertial acceleration $a$ increases and decreases as the gravitational acceleration $g$ decreases. Furthermore, the acceleration $a$ of $M$ produces the conceptual $R$ along with the corresponding variations of binary potential $\delta \varphi_{k}$ \eqref{54} of $\left\lfloor N_{A}\right\rfloor$ bits on a holographic sphere $A$. Each distribution of these bits differs from the BH equilibrium \eqref{56}. Thus, in this case, $N_{1}(t)<N_{A}(t) / 2$ and the mass $M$ alone is a dissipative structure. The dissipativity of this mass, e.g. a rocket, accelerating by itself, can be attributed to the fact that the inertial acceleration can be a result of only two factors: either the rocket accelerates due to gravity or it is accelerated by an engine. In the former case, the gravity of a stellar object can pull the rocket toward its surface (Fig.~\ref{Fig_11}(b)), and the rocket can either escape away or orbit around it (Fig.~\ref{Fig_11}(c), dissipative case). In the latter case (Fig.~\ref{Fig_11}(d)), the dissipativity is provided only by the engine of the rocket. We neglected the situation of the law of conservation of momentum (interaction of rocket with another moving object).

Therefore, the cases depicted in Fig.~\ref{Fig_11}(b-d) are identical in nature. The left sphere displayed in Fig.~\ref{Fig_11} is a gradient sphere, that defines the inertial \eqref{80} or gravitational \eqref{25} potential along with the information capacity $N_{A}$, whereas the right sphere denotes the entropy variation sphere (of the variational potential \eqref{50}) defining $\delta R$ and $N_{1}$. $R$ is an artificial spatial coordinate that enables the introduction of the notion of temperature-generating potentials \eqref{25} and \eqref{80}. Similar to the Rayleigh-Bénard convection, the entropic work must involve a heat transfer defined by Unruh \eqref{38} and Hawking \eqref{39} equations. 

Although the equations of motion in the abstract Newtonian dynamics are mathematically reversible in case the Planck time multiplier $t$ is replaced by $-t$, it is not an observed behavior of the macroscopic objects modeled by these equations. Therefore, although both gravity and inertia appear to be reversible, their origin is entropic.

Any geometric considerations pertaining to the structure of $\mathbb{R}^{3}$ outside the holographic screen are unnecessary. The ancient Greek notion of geometry brings forth an unnecessary burden and axiomatization: \emph{geo-} suggests unobservable extra-instantaneous existence, whereas \emph{-metric} indicates the identity of indiscernibles, an ontological principle and the first axiom of the metric. The identity of indiscernibles principle is invalid owing to the Ugly duckling theorem \cite{36,37} that asserts that any two distinct objects are equally similar. Additionally, the identity of indiscernibles, as the first axiom (vanishing distance from a point to itself) of the metric, is also invalid in the quantum domain \cite{38}.

Geometric considerations are even more unnecessary if one explores the graph of nature with certain intrinsic properties that reflect the $2^{\text {nd}}$ law of thermodynamics, wherein the perceived dimensionality of this graph is induced \cite{3} by the peculiar property of the Euclidean space $\mathbb{R}^{4}$, known as exotic $\mathbb{R}^{4}$. Particularly, the equipotential/holographic spheres or event horizons are defined by the sets of vertices in the graph of nature with the same potential (during instantaneous observation), which attains minimum \eqref{27} for a BH. Based on a geometric perspective, an event horizon is a limiting or fundamental screen with only one spherical geometric, exterior side. We used the term ''holographic sphere'' instead of the term ''equipotential sphere'', considering the latter - which suggests permanent equality of potential - contradicts the notion of a binary message that varies with time.

The anomaly observed in the cosmic microwave background (CMB) radiation aligned with the plane of the Solar System, dubbed the ''axis of evil'' [sic] is considered to be an unsolved problem in physics. A report on the quest for proposed solutions deserves a separate article. They include astrophysical foregrounds (such as Solar system dust \cite{dikarev} or contributions from the Kuiper Belt objects \cite{hansen}), artifacts of faulty data analysis, instrumental systematics, and theoretical/cosmological arguments \cite{copi}. However, the proposed models require severe tuning to reproduce the features of the CMB \cite{copi,kamionkowski,land}.
The author is convinced that this anomaly simply contradicts our intuitive concept of the physical space and could be explained within the framework of the variational \eqref{50}, binary \eqref{54}, and inertial \eqref{80} potentials, and the velocity \eqref{91} and acceleration \eqref{96} relations. Although this matter requires further research, this anomaly may be due to the fact that Ampère’s right-hand grip rule induces some temperature gradient, which is substantially perpendicular to the ecliptic of the Solar System, exhibiting skewness owing to the direction of the Sun’s motion.

\section{Black Holes Quantum Statistics}\label{sec:10}
Planck's law for blackbody spectral radiance expressed in terms of wavelength $\lambda$ is
\begin{equation}\label{129}
B_{\lambda}(\lambda, T)=\frac{2 h c^{2}}{\lambda^{5}} \frac{1}{\exp \left(h c / \lambda k_{B} T\right)-1} .
\end{equation}
For a BH, upon wavelength discretization \eqref{4} and substituting BH temperature \eqref{11}, the above equation can be expressed as
\begin{equation}\label{130}
B_{\lambda}(l, d)=\frac{2 h c^{2}}{l^{5} \ell_{P}^{5}} \frac{1}{\exp \left(4 \pi^{2} d / l\right)-1},
\end{equation}
for wavelength or
\begin{equation}\label{131}
B_{v}(l, d)=\frac{2 h c}{l^{3} \ell_{P}^{3}} \frac{1}{\exp \left(4 \pi^{2} d / l\right)-1},
\end{equation}
for frequency (cf. Appendix \ref{app:13}).

\begin{figure}[ht]
\includegraphics[width=0.5\textwidth]{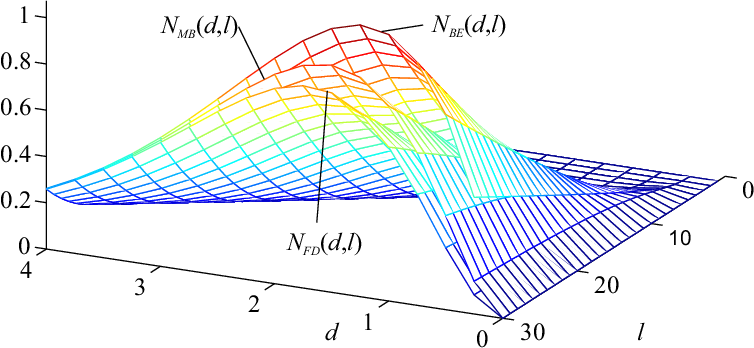}
\caption{\label{Fig_12} Bose-Einstein, Maxwell–Boltzmann, and Fermi-Dirac\\$2$-sphere BH statistics.}
\end{figure}

If a BH Planck area emits (absorbs) a photon (boson), as it alters its state from active $\left(-c^{2}\right)$ to inactive $(0)$ (or vice versa), and an electron (fermion) energy level is decreased (increased) in this interaction, the event horizon can be described using Bose-Einstein (BE) and Fermi-Dirac (FD) statistics with the degeneracy interpreted as the number of Planck areas $\pi d^2$ on the event horizon. No Planck area is distinct and each one is equally capable of emitting/absorbing energy corresponding to the wavelength multiplier $l$. There are certainly no dedicated microwave or X-ray Planck areas on the horizon. Thus, these statistics along with the Maxwell-Boltzmann\footnote{MB statistics lies between BE and FD statistics} (MB) statistics can be described as
\begin{equation}\label{132}
N_{B E}(d, l)=\frac{\pi d^2}{e^{4 \pi^{2} d / l}-1},
\end{equation}
\begin{equation}\label{133}
N_{M B}(d, l)=\frac{\pi d^2}{e^{4 \pi^{2} d / l}}, 
\end{equation}
\begin{equation}\label{134}
N_{F D}(d, l)=\frac{\pi d^2}{e^{4 \pi^{2} d /l}+1},
\end{equation}
as shown in Fig.~\ref{Fig_12}, where $N(d, l)$ represents the average number of bosons, classical \emph{particles}, and fermions, respectively, with wavelength $l$.

\begin{figure}[ht]
\includegraphics[width=0.5\textwidth]{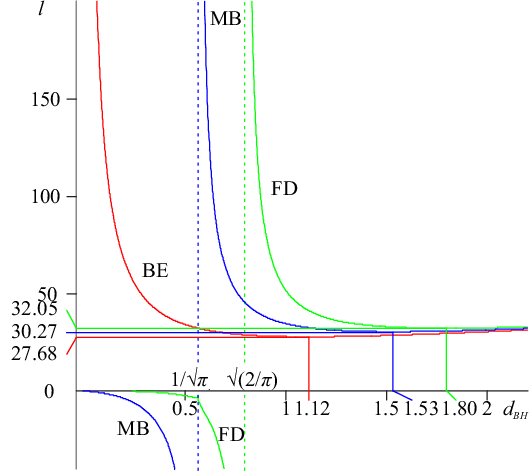}
\caption{\label{Fig_13} Wavelength thresholds for BE (red), MB (blue), and FD
(green) statistics for the $2$-sphere BH as a function of its
diameter multiplier $d_{BH}$ with $N = 1$.}
\end{figure}

To emit/absorb $N>0$ photons or correspondingly increase/decrease $N>0$ electron energy levels on average, $N(d, l)$ must be greater or equal to $N$ and the statistics \eqref{132}-\eqref{134} yield the following wavelength thresholds
\begin{equation}\label{135}
l_{B E}(d, N) \geq \frac{4 \pi^{2} d}{\ln \left(\pi d^{2}+N\right)-\ln (N)} ,
\end{equation}
\begin{equation}\label{136}
l_{M B}(d, N) \geq \frac{4 \pi^{2} d}{\ln \left(\pi d^{2}\right)-\ln (N)} ,
\end{equation}
\begin{equation}\label{137}
l_{F D}(d, N) \geq \frac{4 \pi^{2} d}{\ln \left(\pi d^{2}-N\right)-\ln (N)},
\end{equation}
shown in Fig.~\ref{Fig_13} for $N=1$.

Each threshold is singular at
\begin{equation}\label{138}
d_{B E}^{\sin }=0, \quad d_{M B}^{\sin }=\sqrt{\frac{N}{\pi}}, \quad d_{F D}^{\sin }=\sqrt{\frac{2 N}{\pi}},
\end{equation}
and minimal at
\begin{equation}\label{139}
\begin{split}
&d_{B E}^{\min }=\sqrt{-N\left(\frac{W_{0}\left(-2 e^{-2}\right)+2}{\pi W_{0}\left(-2 e^{-2}\right)}\right)},\\
&d_{M B}^{\min }=e \sqrt{\frac{N}{\pi}}, \quad d_{F D}^{\min }=\sqrt{N \frac{W_{0}\left(2 e^{-2}\right)+2}{\pi W_{0}\left(2 e^{-2}\right)}},
\end{split}
\end{equation}
where $W_{0}(x)$ is the Lambert $W$ function (omega function). Table~\ref{table6} summarizes the minima along with the corresponding minimum wavelengths.

Perhaps, the negative wavelengths for positive diameters can be interpreted in terms of emission/absorption, wherein the positive and negative wavelengths represent the BH absorption and emission, respectively. Consequently, the BE threshold \eqref{135} with no negative wavelengths would illustrate that bosons (photons) cannot escape from a BH. However, a relation between a bit (Planck area) of a binary message \eqref{58} and a wavelength multiplier $l$ remains to be researched.

\begin{figure}[ht]
\includegraphics[width=0.5\textwidth]{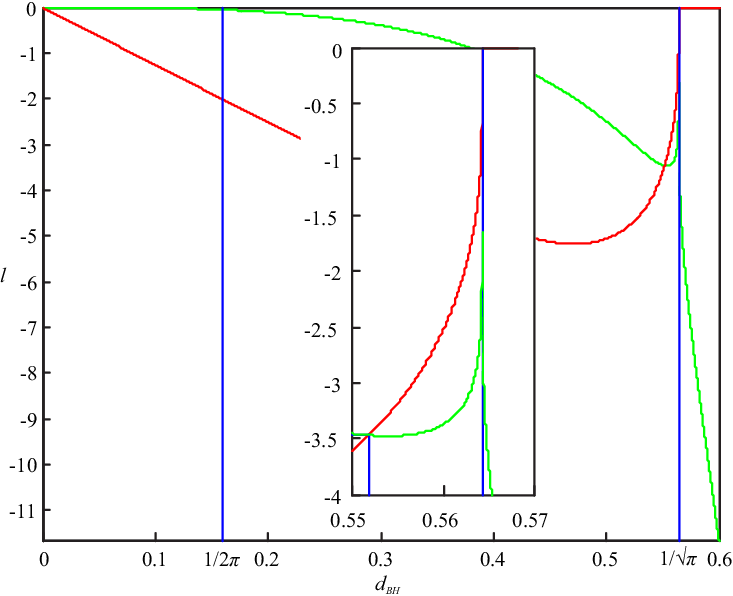}
\caption{\label{Fig_14}  Real (green) and imaginary (red) component
of wavelength threshold for FD statistics as a function of $d_{BH}$
in the vicinity of $1/\sqrt{\pi}$ with $N = 1$.}
\end{figure}

\begin{table}[ht]
\caption{\label{table6} BE, MB, and FD thresholds for $N = 1, 2,\ldots,6.$}
\begin{ruledtabular}
\begin{tabular}{|l|l|l|l|l|l|l|}
$N$ & $d^{\min}_{BE}$ & $l^{\min}_{BE}$ & $d^{\min}_{MB}$ & $l^{\min}_{MB}$ & $d^{\min}_{FD}$ & $l^{\min}_{FD}$ \\
\hline
 1 & $1.1173$ & $27.678$ & $1.5336$ & $30.273$ & $1.8007$ & $32.054$ \\
 2 & $1.5800$ & $39.142$ & $2.1689$ & $42.812$ & $2.5465$ & $45.332$ \\
 3 & $1.9352$ & $47.939$ & $2.6563$ & $52.434$ & $3.1188$ & $55.520$ \\
 4 & $2.2345$ & $55.355$ & $3.0673$ & $60.545$ & $3.6013$ & $64.109$ \\
 5 & $2.4982$ & $61.889$ & $3.4293$ & $67.692$ & $4.0264$ & $71.676$ \\
 6 & $2.7367$ & $67.796$ & $3.7566$ & $74.152$ & $4.4107$ & $78.517$ 
\end{tabular}
\end{ruledtabular}
\end{table}

For $0<d \leq \sqrt{N / \pi}$, the wavelengths $l$ obtained from FD bound \eqref{137} are complex, as shown in Fig.~\ref{Fig_14}. Below this threshold, corresponding to the diameter of a 1-bit BH for $N=1$, BH emits complex wavelengths. After exceeding this diameter, they become \emph{real}, and hence, observable. Presumably, only the negative and real wavelengths in the FD threshold \eqref{137} within the range $\sqrt{N / \pi} < d_{BH} \leq \sqrt{2 N / \pi}$ represent the Hawking radiation of fermions. Interestingly, the real component of the wavelength below $d_{BH}=1 /(2 \pi)$ the Planck temperature threshold vanishes.

As described above, a more detailed description of a BH interaction with the environment certainly requires further research, perhaps considering the imaginary set of Planck units \cite{39}. The degenerated, patternless Bose-Einstein BH blackbody radiation is characterized by zero Gibbs free energy, whereas the departure from degeneracy is responsible for unique spectral characteristics of stars \cite{40}.

\section{Black Hole Information Paradox Solution}\label{sec:11}
To illustrate the black hole information paradox that also applies to neutron stars and white dwarfs emitting patternless blackbody radiation \eqref{129}, we consider two quantum, two-state \emph{particles} (qubits) $A$ and $B$ entangled in one out of four maximally entangled Bell states
\begin{equation}\label{140}
\left|\Phi^{\pm}\right\rangle=\frac{1}{\sqrt{2}}\left(|0\rangle_{A}|0\rangle_{B} \pm|1\rangle_{A}|1\rangle_{B}\right),
\end{equation}
or
\begin{equation}\label{141}
\left|\Psi^{\pm}\right\rangle=\frac{1}{\sqrt{2}}\left(|0\rangle_{A}|1\rangle_{B} \pm|1\rangle_{A}|0\rangle_{B}\right),
\end{equation}
the coherence of which is maintained. If we throw one of these \emph{particles} (say $B$ ) into a BH event horizon, we lose all the information contained in \emph{particle} $B$, as asserted by the no-hiding theorem \cite{1}. However, this information is entangled with the information support of the \emph{particle} $A$ through surjective isometry of the state \eqref{140} or \eqref{141}. Does that mean that qubit $A$ can no longer be measured? Or, if it can be measured, yielding - say - $|0\rangle_{A}$ in a basis $\{|0\rangle, |1\rangle\}$, does that mean that the \emph{particle} $B$ has been earlier measured by the BH event horizon also as $|0\rangle_{B}$ (in the case of $|\Phi^{\pm}\rangle$) or as $|1\rangle_{B}$ (in the case of $|\Psi^{\pm}\rangle$)?

Certainly not. Even if a BH horizon is a quantum system, it cannot stand in the role of an observer \cite{41,42}. Patternless BH horizon contradicts the entanglement between $A$ and $B$ defined by Equations \eqref{140} or \eqref{141}. Any state of the orthonormal basis of the four Bell states in a Hilbert space $\mathbb{C}^{4}$ is a unique pattern among an uncountably infinite number of other possible quantum states. Therefore, BH would destroy the coherence of the entangled state \cite{Teleportation_with_a_Uniformly} transforming it into a separable state
\begin{equation}\label{142}
|\varphi\rangle=\frac{1}{2}\left(|0\rangle_{A}|0\rangle_{B} \pm|0\rangle_{A}|1\rangle_{B} \pm|1\rangle_{A}|0\rangle_{B} \pm|1\rangle_{A}|1\rangle_{B}\right),
\end{equation}
from which the measurement of $A$ will yield an arbitrary, equally probable result. The principle of locality is invalid in the quantum domain \cite{43}, whereas much of the essence of quantum theory already makes itself known in the case of just two nonorthogonal states \cite{30}.

These limits do not hold for the observation of \emph{particle} $B$ on a patterned holographic sphere, as shown in Fig.~\ref{Fig_08}. If $B$ is a photon, observed through an eye of a living organism, its photoisomerization-induced signal transduction cascade will provide a definite result preserving entanglement. In other words, the information obtained from the measured qubit $B$ will penetrate the interior of the observer through a Planck area $\ell_{P}^{2}$ on a holographic sphere, as one bit of classical information. This cannot happen for an interior-less BH shown in Fig.~\ref{Fig_02}(b).

Considering a BH horizon as a quantum system, each BH patternless message \eqref{59} has exactly one orthogonal counterpart (e.g., 0101 is orthogonal to 1010). Thus, the number of orthogonal swaps that BH provides is
\begin{equation}\label{143}
\left(\begin{array}{c}
N_{BH} \\
N_{BH}/2
\end{array}\right) / 2 = \frac{N_{BH}!}{2\left[\left(\frac{N_{BH}}{2}\right)!\right]^2}
\approx \frac{2^{N_{BH}}}{ \sqrt{2 \pi N_{BH}} }=\frac{2^{\pi d_{BH}^2}}{ \pi d_{BH} \sqrt{2}},
\end{equation}
which approaches infinity with $N_{BH}$ approaching infinity, and where Stirling’s approximation extends this formula to odd $N_{BH}$. The Margolus-Levitin theorem and BH mass-energy equivalence \eqref{64} set a bound on a minimum time period required for the BH to transfer from one state to another, orthogonal one, given as
\begin{equation}\label{144n}
\delta t_{M L} \geq \frac{\pi \hbar}{2 E_{BH}}=\frac{2 \pi}{d_{BH}} t_{P},
\end{equation}
which is longer than the Planck time only for $d_{BH} < 2 \pi$, providing another argument for micro BHs instability. Perhaps the Margolus–Levitin bound is not applicable for larger BHs. For 1-bit BH the Margolus–Levitin bound is
\begin{equation}\label{145}
\delta t_{M L} \geq 2 \sqrt{\pi^3}t_{P} \approx 11.137 t_{P},
\end{equation}
which is the minimum time, 1-bit BH needs to swap its Planck area from 0 to 1 [$-c^2$] or vice versa. A relation between Equation \eqref{143} and \eqref{144n} requires further research; certainly, the more orthogonal swaps available, the shorter the swap time.

\section{Conclusions}\label{sec:12}
This study defined BHs as interior-less entities \eqref{13} showing that they emit degenerate, patternless binary messages \eqref{56}, whereas BH diameter $D_{BH}=\ell_{P} /(2 \pi)$ represents the Big Bang conditions.

Micro-BHs are inherently unstable and prone to collapses \cite{20}, and hence, this study delivered further arguments in support of this claim. BH threshold of distinguishability \eqref{10} was found to be a threshold of BH collapsibility \eqref{22}. Temperature uncertainty for one degree of freedom was shown to be equal to the Planck temperature if time uncertainty equals the Planck time \eqref{67}, whereas $\pi$ degrees of freedom enable the recovery \eqref{68} of exact Unruh \eqref{38} and Hawking \eqref{39} temperature equations.

The variational potential \eqref{50} was introduced and shown that it can be expressed as the ratio of the number of the active Planck areas to the information capacity of the holographic sphere \eqref{62}, whereas the entropic work is the product of the test mass and the variational potential \eqref{81}.

Postulated Pythagorean relation \eqref{96} between an observable acceleration $a$ and an unobservable orthogonal acceleration $a_T$ derives \eqref{100} the smallest BH ($d_{BH}=1$) having surface gravity \eqref{82} equaling the Planck acceleration.

Postulated Pythagorean relation \eqref{91} between an observable velocity $v$ and an unobservable orthogonal velocity $v_R$ showed that Lorentz contraction \eqref{93}, \eqref{110} and time dilation \eqref{94}, \eqref{109} formulas can be simplified to become independent of the observer's velocity.

Both velocity \eqref{91} and acceleration \eqref{96} relations introduce significant BH radius equal to $4GM/c^2$, that is twice its Schwarzschild radius ($2GM/c^2$), and exceeds the BH photon sphere radius ($3GM/c^2$) by $GM/c^2$. 

Shannon entropy \eqref{126} of holographic spheres, as a function of $GM/c^2$ multiplier, was derived, showing that there are fewer active Planck areas $N_1$ \emph{outside} the BH horizon than $N_1 = N_A/2$, where $N_A$ is the information capacity of the holographic sphere. 

Pythagorean \eqref{116} and hyperbolic \eqref{117} relations between segments orthogonal and tangential to the holographic sphere revealed $c^2/2$ squared velocity at which de Broglie wavelength (of a matter wave) corresponds to the Compton wavelength of that wave.

Certain dynamics scenarios between the passive mass $M_{BH}$ and the active mass $M$ were discussed.

Bose-Einstein (BE), Maxwell–Boltzmann (MB), and Fermi–Dirac (FD) statistics \eqref{132}–\eqref{134} were derived with the degeneracy interpreted as the number of Planck areas $\pi d^2$ on the event horizon. 

The solution to BH information paradox was proposed \eqref{142} and discussed in the context of the Margolus–Levitin theorem.

A relation between a bit of a binary message \eqref{59} and wavelength multiplier $l$ of blackbody radiation requires further research and may shed new light on the observed ''axis of evil'' cosmological anomaly. An in-depth study of BH interaction with the environment is required.

Perhaps the most striking conclusion of this study is that every dissipative structure is a sphere in nonequilibrium thermodynamic condition, provided with an interior. Interiors of living cells were exploited by biological evolution, probably beginning with coacervates. An open question is why would a hurricane need an interior?

\begin{acknowledgments}
I truly thank my wife and her mother for their support. I would like to thank Mirek for the insightful discussions, continuous encouragement during these three years of research, and in particular for noting that the concept of a bit is inadequate to be meaningfully assigned to the notion of the potential, for noting that BHs invalidate the Jordan–Brouwer separation theorem, and for assisting while solving the \emph{sign riddle}. I would like to thank my godson Wawrzyniec for meticulously spotting typographical errors and missing commas, for his report on the ''axis of evil'' state of the art, and for a valuable \emph{external perspective} from \emph{outside} of my holographic sphere. I also thank Piotr for the helpful discussions and for finding certain logical loopholes in the paper, along with noting that the effect of the ''axis of evil'' could be due to Ampère's right-hand grip rule. Finally, I thank Mariola for introducing order and discipline in our \emph{Fellowship of the Ring}, without which this study would certainly not have been completed.
\end{acknowledgments}

\appendix*

\section{}
\subsection{The alternative form of Hawking temperature}\label{app:1}
Substituting BH surface gravity \eqref{82} into Hawking blackbody-radiation equation \eqref{39} and using \eqref{3} yields
\begin{equation}\label{144}
T_{BH}=\frac{\hbar}{2 \pi c k_{B}} \frac{c^2}{D_{BH}} = \frac{\hbar c}{2 \pi k_{B} d  \ell_P} =\frac{T_P}{2 \pi d}.
\end{equation}

\subsection{The alternative form of Planck-Einstein (Compton) relation}\label{app:2}
The Planck relation can be expressed in terms of the Planck temperature, using \eqref{4}, as   
\begin{equation}\label{145n}
\begin{split}
E=&h \nu=h \frac{c}{\lambda}=\frac{h c}{l \ell_P}=\frac{2 \pi \hbar c}{l} \sqrt{\frac{c^3}{\hbar G}}= \\
=&\frac{2 \pi}{l} \frac{k_B}{k_B} \sqrt{\frac{\hbar^2 c^2 c^3}{\hbar G}}=\frac{2 \pi}{l} k_B T_P.
\end{split}
\end{equation}

\subsection{BH diameter fluctuations}\label{app:3}
After the absorption/emission of a wavelength $l$, the 2-sphere BH diameter defined by its mass increases(+)/decreases(-) owing to the Compton mass $M=h /(c \lambda)$ of this wavelength. Therefore, using \eqref{3}
\begin{equation}\label{146}
\begin{split}
D_{BH}^{A/E}=&D_{BH} \pm \delta D=d \ell_P \pm \frac{4 G}{c^2} \frac{2 \pi \hbar}{c l \ell_P} \\
=&d \ell_P \pm \frac{8 \pi \hbar G}{l c^3 \ell_P}=d \ell_P \pm \frac{8 \pi \ell_P}{l}=\left(d \pm \frac{8 \pi}{l}\right) \ell_P
\end{split}
\end{equation}
Accordingly, the BH area also increases/decreases
\begin{equation}\label{147}
A_{BH}^{A/E}=\pi \left( D_{BH}^{A/E} \right)^2 =\pi\left((d \pm \frac{8 \pi}{l}\right)^2 \ell_P^2.
\end{equation}
If $l=d / 2$ \cite{18,19} then
\begin{equation}\label{148}
d_{k+1}^{A/E}=\pm \sqrt{256 \pi^2 \frac{1}{d_k^2} \pm 32 \pi+d_k^2}.
\end{equation}

\subsection{BH diameter after absorption/emission of a wavelength}\label{app:4}
BH diameter \eqref{20} after absorption (+) or emission (-) of a wavelength $l$ in dimensions $n=2,3,4,5$.
\begin{table}[h]
\begin{ruledtabular}
\begin{tabular}{|l|l|}
$n$ & $\left(d_{k+1}^{A / E}\right)^{n-1}$ \\
\hline
 2 & $d_k \pm 8 \pi / l$ \\
 3 & $d_k^2 \pm 16 \pi d_k / l+64 \pi^2 / l^2$ \\
 4 & $d_k^3 \pm 24 \pi d_k^2 / l+192 \pi^2 d_k / l^2 \pm 512 \pi^3 / l^3$ \\
 5 & $d_k^4 \pm 32 \pi d_k^3 / l+384 \pi^2 d_k^2 / l^2 \pm 2048 \pi^3 d_k / l^3+4096 \pi^4 / l^4$
\end{tabular}
\end{ruledtabular}
\end{table}

\subsection{Negative and imaginary unit lengths}\label{app:5}
There are four possibilities for the volume of unit $n$ cubic element $d^n$ in $n$-dimensional space in dependence on what we take as the unit length $d=\{1,-1, i,-i\}$. Thus for $n \in \mathbb{C}$, and for $a \in \mathbb{R}$, $a>0$
\begin{equation}\label{149}
a^n=a^{\operatorname{Re}(n)} \{\cos \left[ \operatorname{Im} (n) \ln (a) \right] + i \sin \left[ \operatorname{Im} (n) \ln (a) \right] \},
\end{equation}
(which equals one $\forall n \in \mathbb{C}$ if $a=1$), for $a \in \mathbb{R}$, $a \le 0$
\begin{equation}\label{150}
a^n=|a|^n i^{2n} = |a|^n \begin{cases}1 & 2n = 4k, k \in \mathbb{Z} \\ -1 & 2n=4k+2 \\ \mp i & 2n=4k \mp 1 \\ \mathbb{C} & \text{otherwise}\end{cases},
\end{equation}
whereas for imaginary $a = b i$, $b \in \mathbb{R}$
\begin{equation}\label{151}
a^n= |b|^n (\pm i)^n = |b|^n \begin{cases}1 & n=4k, k \in \mathbb{Z} \\-1 & n=4k+2\\ \mp i & n=4k \mp 1\\ \mathbb{C} & \text{otherwise}\end{cases}.
\end{equation}

The unit length $d$ is a scaling factor of a relation between the vertices of the graph of nature. Generally, it can be real (positive or negative) or imaginary (positive or negative), and this list may not be exhaustive. Any vertex beyond the countably infinite number of vertices can exhibit any relation with other vertices. However, only a real and positive unit length \eqref{149} yields a real and positive unit $n$-cubic element $d^n$ for any complex dimension $n$. We note that in passing that the law of cosines, which generalizes the Pythagorean theorem, does not introduce imaginary lengths, only negative ones.

\subsection{BH entropy \eqref{44} from Verlinde's entropy variation \eqref{46}}\label{app:6}
\begin{equation}\label{153}
\begin{split}
\delta S &=2 \pi k_B \frac{M c}{\hbar} \delta R \myeq -2 \pi k_B \frac{M_{BH} c}{\hbar} R_{BH}=\\
&=-2 \pi k_B \frac{c}{\hbar} \frac{R_{BH} c^2}{2 G} R_{BH}=-k_B \frac{4}{4}\frac{ \pi R_{BH}^2}{\ell_P^2}= \\
&=-\frac{1}{4} k_B N_{BH},
\end{split}
\end{equation}
where $R$ to $\delta R$ relation \eqref{87} was used, so the sign is reversed.

\subsection{BH temperature vs. BH energy/work}\label{app:7}
Considering BH having the Planck temperature and the Planck energy yields, using Equations \eqref{11} and \eqref{64}
\begin{equation}\label{154}
\frac{T_{BH}}{T_P}=\frac{1}{2 \pi d_{BH}} \cap  \frac{E_{BH}}{E_P}=\frac{d_{BH}}{4} \Rightarrow d_{BH}=\pm \sqrt{\frac{2}{\pi}},
\end{equation}
that is $d_{BH}$ of 2-bit BH and the $1^{\text {st }}$ singularity \eqref{138} of the FD statistics.

Considering BH having the Planck temperature and doing BH entropic work \eqref{63} equaling the Planck energy yields
\begin{equation}\label{155}
\frac{T_{BH}}{T_P}=\frac{1}{2 \pi d_{BH}} \cap \frac{W_{BH}}{E_P}=\frac{d_{BH}}{8} \Rightarrow d_{BH}=\pm \sqrt{\frac{4}{\pi}},
\end{equation}
that is $d_{BH}$ of 4-bit BH, one unit of BH entropy \eqref{44}, and the $2^{\text {nd }}$ singularity \eqref{138} of the FD statistics.

Comparing BH mass-energy equivalence \eqref{64} with the Landauer limit yields
\begin{equation}\label{156}
\begin{split}
&M_{BH} c^2=T_{BH} k_B \ln \left(2\right) \\
&\frac{d_{BH} \ell_P c^2}{4 G} c^2=\frac{T_P}{2 \pi d_{BH}} k_B \ln (2) \\
&\frac{d_{BH} c^4}{4 G} \sqrt{\frac{\hbar G}{c^3}}=\sqrt{\frac{\hbar c^5}{G k_B^2}} \frac{k_B \ln (2)}{2 \pi d_{BH}} \\
&\frac{d_{BH}}{2}=\frac{\ln (2)}{\pi d_{BH}} \quad d_{BH}=\sqrt{2} \sqrt{\frac{\ln (2)}{\pi}} \approx 0.6643 \\
&N_{BH}=\pi d_{BH}^2=2 \ln (2) \approx 1.3863.
\end{split}
\end{equation}

Comparing BH entropic work \eqref{63} with the Landauer limit, or equivalently, the BH entropy with the Landauer limit entropy yields
\begin{equation}\label{157}
\begin{split}
&\frac{1}{2} M_{BH} c^2=T_{BH} k_B \ln (2) \quad \text { or } \\
&\frac{1}{4} k_B N_{BH}=k_B \ln (2) \\
&N_{BH}=4 \ln (2)=\pi d_{BH}^2 \approx 2.7725 \\
&d_{BH}=2 \sqrt{\frac{\ln (2)}{\pi}} \approx 0.9394.
\end{split}
\end{equation}

\subsection{Variational potential \eqref{78} on a BH horizon}\label{app:8}
\begin{equation}\label{158}
\frac{4 \pi c^2}{N_{BH}} \frac{R_{BH}}{\lambdabar_M}=-\frac{c^2}{2},
\end{equation}
\begin{equation}\label{159}
\frac{4 \pi c^2}{N_{BH}} \frac{R_{BH}}{\lambdabar_M}=\frac{4 \pi c^2}{4 \pi R_{BH}^2 / \ell_P^2} \frac{R_{BH}}{\lambdabar_M}=-\frac{c^2}{2},
\end{equation}
\begin{equation}\label{160}
\lambdabar_M=-\frac{2 \ell_P^2}{R_{BH}}=-\lambdabar_{BH}.
\end{equation}
Furthermore,
\begin{equation}\label{161}
N_{BH}=\frac{4 \pi R_{BH}^2}{\ell_P^2} \Leftrightarrow R_{BH}=\pm \sqrt{\frac{N_{BH}}{4 \pi}} \ell_P,
\end{equation}
so
\begin{equation}\label{162}
\lambda_{BH}=\pm 2 \ell_P \sqrt{\frac{4 \pi}{N_{BH}}} \Leftrightarrow \lambda_M=\mp 2 \ell_P \sqrt{\frac{4 \pi}{N_{BH}}} .
\end{equation}
\subsection{Planck units' derivation}\label{app:9}
We begin with the measured speed of light $c$ and define it to be a quotient of some length unit $\left(\ell_P\right)$ and some time unit $\left(t_P\right): c=\ell_P / t_P$. Thereafter, we define some unit of acceleration $\left(a_P\right)$ as the time unit required to achieve the speed of light: $a_P=c / t_P$. Subsequently, we assume some inertial force equal to gravitational force between two such mass units $\left(m_P\right)$ distanced by the length unit
\begin{equation}\label{163}
m_P a_P=G \frac{m_P m_P}{\ell_P^2} .
\end{equation}
Ultimately, a Compton wavelength or the Planck relation (they are equivalents) for the mass unit is required, which we express as
\begin{equation}\label{164}
m_P=\frac{h}{c \lambda} \Leftrightarrow m_P c^2=h v=h \frac{c}{\lambda} .
\end{equation}
Next, we consider the reduced $\hbar = h/(2 \pi)$ and presume the wavelength to be $\lambda=2 \pi \ell_P$. The remaining derivation is posed as an exercise for the reader.

The derivation above is probably known. However, it has not been found by the author in the state of the art. 

Overall, most likely the simplest derivation of Planck units, based on HUP \eqref{65}, BH mass-energy equivalence \eqref{64}, and the Schwarzschild radius \eqref{26} is presented in \cite{31}.

\subsection{Hawking temperature from the equipartition theorem}\label{app:10}
If the BH energy is uniformly divided over $N_{BH}$ bits on the horizon, the temperature can be determined using BH mass-energy equivalence \eqref{64} and the equipartition theorem for a single degree of freedom (one bit) multiplied by the number of bits $N_{BH}$
\begin{equation}\label{166}
M_{BH}c^2=\frac{1}{2} k_B N_{BH} T_{BH},
\end{equation}
from which relation we can recover the Hawking temperature \eqref{39}, \eqref{11} as
\begin{equation}\label{167}
\begin{split}
T_{BH}=&\frac{2 M_{BH} c^2}{k_B N_{BH}} = \frac{2c^2}{k_B} \frac{D_{BH} c^2}{4 G} \frac{\ell_P^2}{\pi D_{BH}^2} =\frac{c^4 \ell_P^2}{2 \pi d_{BH} k_B G \ell_P}\\
&=\frac{c^4}{2 \pi d_{BH} k_B G } \sqrt{\frac{\hbar G}{c^3}}=\frac{1}{2 \pi d_{BH}} \sqrt{\frac{\hbar G c^8}{k_B^2 G^2 c^3}}\\
&=\frac{1}{2 \pi d_{BH}}\sqrt{\frac{\hbar c^5}{G k_B^2}}=\frac{T_{P}}{2 \pi d_{BH}}.
\end{split}
\end{equation}

\subsection{Potential in negative dimensions}\label{app:11}
\begin{figure}[ht]
\includegraphics[width=0.5\textwidth]{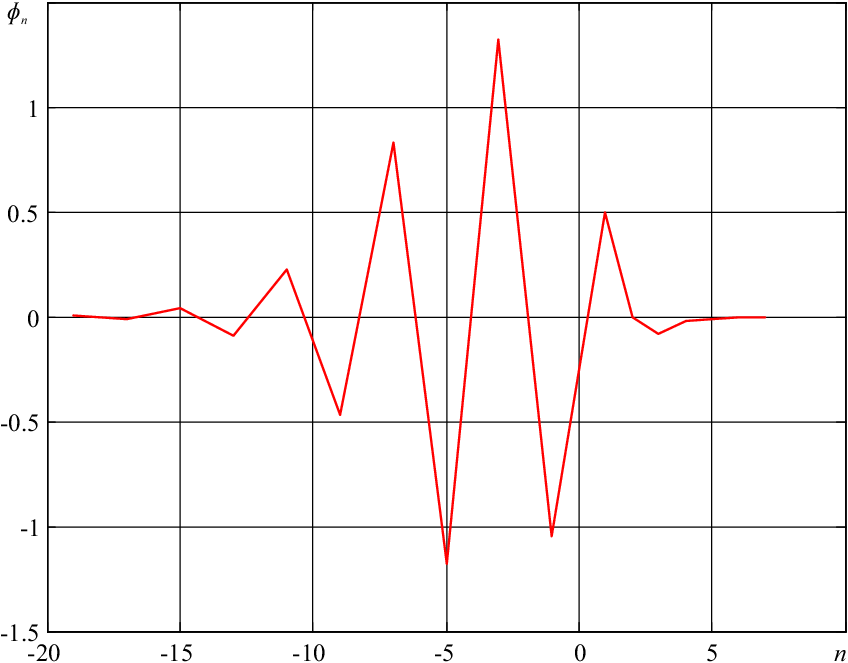}
\caption{\label{Fig_15}  Potential $\phi_n$ in dimensions $n=-20,\ldots,10$, $R=1$.}
\end{figure}

If a spherically symmetric fundamental solution (potential) $\phi$ of Laplace's equation and the general Poisson's equation exists in $\mathbb{R}^n$, it must be in the form of \cite{44, 45}
\begin{equation}\label{168}
\phi_n=\left\{\begin{array}{ll}
\frac{R^{2-n}}{(2-n) n V_n} & \text { for } n \neq 2 \\
\frac{1}{2 \pi} \log (R) & \text { for } n=2
\end{array},\right.
\end{equation}
where $R \neq 0$ and $nV_n$ represents the surface area of the unit $n$-ball (having volume $V_n$). This can be evaluated in the negative dimensions, as shown in Fig.~\ref{Fig_15} for $R=1$.

Further research is required to evaluate the possible relation of the logarithmic potential \eqref{168} in $\mathbb{R}^2$ with Shannon information entropy \eqref{60} and Landauer bound \eqref{61}.

\subsection{Lorentz contraction with velocity relation}\label{app:12}
Using observable and unobservable velocities \eqref{90}, \eqref{104}, and time periods relation \eqref{105}, the velocity relation \eqref{91} becomes
\begin{equation}\label{169}
\frac{\delta L^2}{\delta t^2}-\frac{\delta R^2}{\delta t^2}=c^2 .
\end{equation}
Furthermore, squared Lorentz contraction \eqref{93} can be expressed as
\begin{equation}\label{170}
\delta L^2=-\frac{\delta L_0^2}{c^2} \frac{\delta R^2}{\delta t^2}.
\end{equation}
We substitute $c^2 \delta t^2$ from Equation \eqref{170} into the first Equation \eqref{169} to reduce it to a form that is time-independent.

\subsection{Planck's law}\label{app:13}
We set
\begin{equation}\label{171}
A_\lambda \doteq \frac{2 h}{\ell_P^5} c^2, \quad A_v \doteq \frac{2 h}{\ell_P^3} c^2,
\end{equation}
in Planck's laws for BH spectral radiance \eqref{130}, \eqref{131}. Their derivatives with respect to $l$ are (derivatives with respect to $d$ are monotonic)
\begin{equation}\label{172}
\begin{split}
&\frac{\partial B_\lambda(l, d)}{\partial l}= \\
&\frac{4 A_\lambda d \pi^2 \exp \left(4 \pi^2 d / l\right)}{l^7\left(\exp \left(4 \pi^2 d / l\right)-1\right)^2}-\frac{5 A_\lambda}{l^6\left(\exp \left(4 \pi^2 d / l\right)-1\right)},
\end{split}
\end{equation}
\begin{equation}\label{173}
\begin{split}
&\frac{\partial B_v(l, d)}{\partial l}= \\
&\frac{4 A_v d \pi^2 \exp \left(4 \pi^2 d / l\right)}{l^5\left(\exp \left(4 \pi^2 d / l\right)-1\right)^2}-\frac{3 A_v}{l^4\left(\exp \left(4 \pi^2 d / l\right)-1\right)}.
\end{split}
\end{equation}

From Equations \eqref{172} and \eqref{173}, we can determine the wavelength multiplier $l$ maximizing spectral radiance at a given BH diameter $d$ in terms of the maximum radiation wavelength $\lambda$
\begin{equation}\label{174}
\begin{split}
d_\lambda^{\max }=&\frac{l}{4 \pi^2}\left(W_0\left(-5 e^{-5}\right)+5\right)= \\
=&0.125767812719937 l,
\end{split}
\end{equation}
\begin{equation}\label{175}
l_\lambda^{\max }=7.951159986 d,
\end{equation}
or of the maximum radiation frequency $v$
\begin{equation}\label{176}
\begin{split}
d_v^{\max }=&\frac{l}{4 \pi^2}\left(W_0\left(-3 e^{-3}\right)+3\right)= \\
=&0.071467894189626 l,
\end{split}
\end{equation}
\begin{equation}\label{177}
l_v^{\max }=13.99229698 d,
\end{equation}
where $W_0(x)$ is the Lambert $W$ function.

It can be seen that the maximum emitted wavelength of BH in the visible radiation range ($\lambda=1~\mu m$, $l=\num{6.19E28}$) corresponds to a diameter $d_{BH}=\num{7.78E27}$, $D_{BH}=\num{1.26E-7}$ m, whereas the maximum emitted wavelength of Sagittarius A$^*$ ($D_{BH} \approx \num{2.45E10}$ m) amounts to $\lambda=\num{1.95E11}$ m, which is much larger than the largest EMR wavelength ever detected ($\approx 0.1$ Hz).

\subsection{Certain properties of the binary Shannon entropy \eqref{126}\\of holographic spheres}\label{app:14}
\begin{equation}\label{178}
H(k)=\log (k)-\frac{k-1}{k} \log (k-1) .
\end{equation}
The $1^{\text {st }}$ derivative is
\begin{equation}\label{179}
\frac{d H}{d k}=-\frac{\log (k-1)}{k^2},
\end{equation}
yielding the maximum of $H(k)$ at $k=2$. The $2^{\text {nd }}$ derivative
\begin{equation}\label{180}
\frac{d^2 H}{d k^2}=\frac{2(k-1) \log (k-1)-k}{k^3(k-1)},
\end{equation}
yields the inflection point of $H(k)$ at
\begin{equation}\label{181}
k=\frac{1}{2 W_0\left(\frac{1}{2} e^{-\frac{1}{2}}\right)}+1 \approx 3.0935.
\end{equation}
The integral is
\begin{equation}\label{182}
\begin{split}
\int H d k &=(\log (k-1)+k) \log (k) \\
&-(k-1) \log (k-1)+\mathrm{Li}_2(1-k),
\end{split}
\end{equation}
where $\mathrm{Li}_2(z)$ denotes dilogarithm.

Furthermore, $H(k)=H\left(\frac{k}{k-1}\right)$.

\subsection{The volume of an $n$-ball in complex dimensions}\label{app:15}
The volume of an $n$-ball ($B$) is known to be
\begin{equation}\label{183}
V_n(R)_B=\frac{\pi^{n/2}}{\Gamma(n/2+1)}R^n,
\end{equation}
where $\Gamma(\mathbb{C}\rightarrow\mathbb{C})$ is the Euler's gamma function and $R$ is the $n$-ball radius. The gamma function is defined for all complex numbers excluding non-positive integers. Thus, for $n = a + ib$, where if $a \in \mathbb{Z}$, $a > 0$,
\begin{equation}\label{184}
\begin{split}
\pi^{n/2} =& \pi^{(a+ib)/2} = \\ &=\pi^{a/2}\left[\cos\left(\frac{b}{2}\ln(\pi)\right) + i\sin\left(\frac{b}{2}\ln(\pi)\right) \right],
\end{split}
\end{equation}
\begin{equation}\label{185}
R^n = R^{a+ib} = R^a\left[\cos\left(b\ln(R)\right) + i\sin\left(b\ln(R)\right) \right],
\end{equation}
the volume becomes
\begin{equation}\label{186}
V_n(R)_B=\frac{\pi^{a/2}R^a\left[\cos\left(b\ln(R\sqrt{\pi})\right) + i\sin\left(b\ln(R\sqrt{\pi})\right) \right]}{\Gamma\left(\frac{a+ib}{2}+1\right)},
\end{equation}
and the surface becomes
\begin{equation}\label{187}
\begin{split}
S_n(R)_B=&\frac{n}{R}V_n(R)_B=\pi^{a/2}R^{a-1}(a+ib)\\
&\frac{\left[\cos\left(b\ln(R\sqrt{\pi})\right) + i\sin\left(b\ln(R\sqrt{\pi})\right) \right]}{\Gamma\left(\frac{a+ib}{2}+1\right)}.
\end{split}
\end{equation}

The anti-symmetry of the imaginary part of the volume \eqref{186} and the surface \eqref{187}, in a way, establishes the arrow of time and is independent of $\operatorname{Re}(n)$ for $\operatorname{Im}(n) = 0$.

\nocite{*}

\bibliographystyle{ieeetr}
\bibliography{apssamp}

\end{document}